\documentclass[draftclsnofoot,onecolumn]{IEEEtran}
\usepackage{amsmath,amsfonts,amssymb}
\usepackage[thmmarks,amsmath]{ntheorem}
\usepackage{cases}
\usepackage{graphicx}
\usepackage{fixfoot}
\usepackage{stfloats,balance}
\usepackage{epsfig}
\usepackage[noend]{algpseudocode}
\usepackage{algorithmicx,algorithm}
\usepackage{cite}
\usepackage{xcolor}
\usepackage{subeqnarray}
\usepackage{textcomp}
\usepackage{bm}
\usepackage{multicol}  
\usepackage{multirow} 
\usepackage{booktabs}
\theoremseparator{:}
\theoremheaderfont{\itshape}

\newtheorem{theorem}{Theorem}

\newtheorem{remark}{Remark}
\newtheorem{property}{Property}

\allowdisplaybreaks[3]
\def\BibTeX{{\rm B\kern-.05em{\sc i\kern-.025em b}\kern-.08em
    T\kern-.1667em\lower.7ex\hbox{E}\kern-.125emX}}

\allowdisplaybreaks[4] 

\begin{document}
\title{MEC-Empowered Non-Terrestrial Network for 6G Wide-Area Time-Sensitive \\ Internet of Things}

 \author{Chengxiao~Liu, Wei~Feng, Xiaoming~Tao and Ning~Ge

\thanks{C. Liu, W. Feng, X. Tao and N. Ge are with the Department of Electronic Engineering, Tsinghua University, Beijing 100084, China, and with the Beijing National Research Center for Information Science and Technology, Tsinghua University, Beijing 100084, China, and also with the Beijing Innovation Center for Future Chip, Beijing 100084, China (email: lcx17@mails.tsinghua.edu.cn, fengwei@tsinghua.edu.cn, taoxm@mail.tsinghua.edu.cn, gening@tsinghua.edu.cn).}

}
\maketitle
\begin{abstract}
In the upcoming sixth-generation (6G) era, the demand for constructing a wide-area time-sensitive Internet of Things (IoT) keeps increasing. As conventional cellular technologies are hard to be directly used for wide-area time-sensitive IoT, it is beneficial to use non-terrestrial infrastructures including satellites and unmanned aerial vehicles (UAVs), where a non-terrestrial network (NTN) can be built under the cell-free architecture. Driven by the time-sensitive requirements and uneven distribution of IoT devices, the NTN is required to be empowered by mobile edge computing (MEC) while providing oasis-oriented on-demand coverage for the devices. Nevertheless, communication and MEC systems are coupled with each other under the influence of complex propagation environment in the MEC-empowered NTN, which makes it hard to orchestrate the resources. In this paper, we propose a process-oriented framework to design the communication and MEC systems in a time-division manner. Under this framework, the large-scale channel state information (CSI) is used to characterize the complex propagation environment with an affordable cost, where a non-convex latency minimization problem is formulated. After that, the approximated problem is given and it can be decomposed into subproblems. These subproblems are further solved in an iterative way. Simulation results demonstrate the superiority of the proposed process-oriented scheme over other algorithms. These results also indicate that the payload deployments of UAVs should be appropriately predesigned to improve the efficiency of resource use. Furthermore, the results imply that it is advantageous to integrate NTN with MEC for wide-area time-sensitive IoT.
\end{abstract}

\section{Introduction}
\par{
In future sixth-generation (6G) networks, the concentration of cutting-edge technologies will change from humans to intelligent machines \cite{Saarnisaari2020}. Different from human beings, these machines are usually unevenly distributed in remote areas \cite{Saarnisaari2020} and built for accomplishing time-sensitive tasks \cite{Network2030}. This scenario increases the demands for constructing a wide-area time-sensitive Internet of Things (IoT) in the upcoming 6G era \cite{Saarnisaari2020,Network2030}.
}
\par{
However, terrestrial infrastructures are hard to be deployed in remote areas \cite{Wei2021,Li2020}, which indicates that terrestrial cellular networks have blind sides in terms of coverage ability \cite{Onireti2016}. As a result, it is difficult to serve intelligent machines using conventional fourth-generation (4G) and fifth-generation (5G) technologies. To overcome this obstacle, it is beneficial to use non-terrestrial infrastructures including satellites and unmanned aerial vehicles (UAVs) for wide-area time-sensitive IoT, where a non-terrestrial network (NTN) can be built. Specifically, to accommodate the uneven distribution of machines, such NTN is needed to provide oasis-oriented on-demand coverage for machines, which indicates that the NTN should be designed under a cell-free architecture \cite{Liu2020}. In addition, driven by the time-sensitive requirements of machines, data from machines must be processed by the NTN as quickly as possible. For this purpose, SatCom-on-the-move antennas can be carried on UAVs to build high-speed links between satellites and UAVs \cite{Zhao2018}, while edge servers can also be carried on UAVs for rapidly processing data with mobile edge computing (MEC) \cite{Cheng2019}. Thus, an MEC-empowered NTN is required to be constructed under the cell-free architecture. Nevertheless, communications and MEC are coupled with each other in the NTN with complex propagation environment, which implies that new difficulties will arise. First, due to the complex propagation environment in the MEC-empowered NTN, it is challenging to realize oasis-oriented on-demand coverage under the cell-free architecture \cite{Liu2020}. Additionally, as communication and MEC systems are coupled with each other in this NTN, jointly orchestrating the resources is rather complicated \cite{Cheng2019}. As a result, we investigate the design of an MEC-empowered NTN for wide-area time-sensitive IoT in this work, as this remains an open question to our knowledge.
}

\subsection{Related Works}
\par{
For wide-area IoT, the Narrow-Band IoT (NB-IoT) is an enabling technique that was designed under conventional cellular architecture \cite{Raza2017}, while the Long Range Radio (LoRa) technique was proposed to further expand network coverage \cite{Centenaro2016}. In addition, to serve time-sensitive machines, the design of time-sensitive networks (TSNs) has attracted extensive attention worldwide \cite{LoBello2019}, where industrial automation is an essential application scenario \cite{LoBello2019,Liang2019,Luvisotto2019}. Bello {\em et al.} gave an overview of the applicability of TSN to various industrial systems in \cite{LoBello2019}. Liang {\em et al.} presented a comprehensive survey on Wireless networks for Industrial Automation–Factory Automation (WIA-FA) technique and its applications in \cite{Liang2019}. Luvisotto {\em et al.} evaluated the feasibility of Wireless High Performance (WirelessHP) technology for industrial wireless networks in \cite{Luvisotto2019}. These researches promoted the standardization of 5G ultra-reliable low-latency communication (URLLC) \cite{3GPPURLLC} and industrial IoT \cite{3GPPIIoT} conducted by 3rd Generation Partnership Project (3GPP). }
\par{
Due to the coverage holes of terrestrial cellular networks, NTN may become a promising technique for 6G networks, where the standardization of NTN has been launched in 3GPP Release~16 \cite{3GPPNTN}. In the future, how to design an NTN for supporting a wide-area time-sensitive IoT would be discussed in 3GPP Release 17 \cite{Ghosh2019}. In existing studies related to NTNs, satellite-enabled IoT has been widely discussed, as it can provide ubiquitous coverage for wide-area IoT \cite{Sanctis2016, Cioni2018, Zhen2019}. Sanctis {\em et al.} investigated the protocols and architectures of satellite-based Internet for remote things in \cite{Sanctis2016}. Cioni {\em et al.} studied the opportunities and challenges of satellite-enabled massive machine-type communication (MMTC) in \cite{Cioni2018}. Zhen  {\em et al.} took a step further to propose an optimal preamble design method in \cite{Zhen2019}, which can adapt to the group-based random access pattern for satellite-based MMTC. However, satellite-enabled IoT systems suffer from high latency and low efficiency \cite{Cioni2018,Sanctis2016,Zhen2019}, which make it difficult to meet the requirements of intelligent machines \cite{Saarnisaari2020}.
}
\par{
In addition, the UAVs have potential in providing on-demand services for wide-area time-sensitive IoT \cite{Zhang2019,Chakareski2019,Ranjha2020}. In \cite{Zhang2019}, a low-latency routing algorithm was proposed for UAV-enabled IoT, which was designed under a layered network architecture with UAV swarm. The author of \cite{Chakareski2019} proposed the design of a UAV-enabled IoT-oriented network to support real-time remote virtual reality. In \cite{Ranjha2020}, the uplink (UL) power of IoT devices was optimized to design a UAV-assisted URLLC network. To further improve latency performance, UAV-enabled IoT is integrated with MEC \cite{Islambouli2019,Zhang2020Latency,Tan2020,Tun2020, Wang2020}. In \cite{Islambouli2019}, the 3D deployment of UAVs was optimized to support time-sensitive IoT, where UAVs are mounted as cloudlets. The authors of \cite{Zhang2020Latency} minimized the average latency of users in UAV-aided MEC networks. In \cite{Tan2020}, trajectories of UAVs were optimized for smart IoT community, where an augmented reality-based use case was discussed. The authors of \cite{Tun2020} proposed an energy-efficient multi-domain resource allocation scheme while considering stringent latency requirements. In \cite{Wang2020}, an online UAV-mounted edge server dispatching scheme was proposed, where latency fairness among users was guaranteed with efficient resource utilization.  Nevertheless, the UAV-enabled network usually lacks persistence and stability \cite{Zeng2019}, which is an inevitable limitation for wide-area time-sensitive IoT. 
}
\par{
In view of previous works, it is advantageous to jointly use satellites and UAVs with MEC for wide-area time-sensitive IoT \cite{Liu2020Task, Cheng2019, Cao2019}. Liu {\em et al.} presented a task-oriented intelligent architecture for IoT-oriented space-aerial-ground-aqua integrated networks in \cite{Liu2020Task}. Cheng {\em et al.} investigated the joint design of computing resource allocation and task offloading strategy for IoT-oriented space-aerial-ground integrated networks in \cite{Cheng2019}, where stringent latency constraints were regarded and a learning-based approach was proposed. Cao {\em et al.} discussed the coupling of trajectory design and task offloading strategy in an integrated satellite-UAV network under wind influence in \cite{Cao2019}. Despite all of these achievements, when an NTN is integrated with MEC under the cell-free architecture, it will encounter new challenges. First, due to the complex propagation environment, channel state information (CSI) is hard to be perfectly acquired by the NTN, which indicates that it is complicated to provide oasis-oriented on-demand coverage for machines under the cell-free architecture. Furthermore, it is challenging to orchestrate the resources in the MEC-empowered NTN because communication and MEC systems are coupled with each other. In our previous works \cite{Liu2020}, we discussed the cell-free coverage patterns of integrated satellite-UAV networks. In this work, we go a step further to investigate the design of an MEC-empowered NTN for wide-area time-sensitive IoT. The relationships between existing technologies and the specific research area that we focus on is summarized in Table~I.
}

\begin{table} 
	\centering  
	\caption{Existing technologies and our concentrations}  
	\begin{tabular}{c|c|c|c}
		\toprule[1pt]
	Specific area & Technology & Coverage pattern & Latency requirement \\
		\hline
	\multirow{3}{*}{Wide-area IoT}	& NB-IoT \cite{Raza2017} & Cellular-based & \multirow{3}{*}{ Large latency allowed} \\
		\cline{2-3}
		& LoRa \cite{Centenaro2016} & Expanded cellular &  \\
		\cline{2-3}
		& Satellite-based IoT \cite{Zhang2019,Chakareski2019,Ranjha2020} & Ubiquitous & \\
		\hline
		\multirow{4}{*}{TSN} & WIA-FA \cite{Liang2019} & \multirow{2}{*}{Indoor} & \multirow{6}{*}{ Sensitive to latency} \\
		\cline{2-2}
		& WirelessHP \cite{Luvisotto2019} &  &  \\
		\cline{2-3}
		 & 3GPP 5G URLLC \cite{3GPPURLLC} & \multirow{2}{*}{Cellular-based} & \\
		\cline{2-2}
		 & 3GPP Industrial IoT \cite{3GPPIIoT} &  & \\ 
		\cline{1-3}
		\multirow{2}{*}{\shortstack{Wide-area time\\ -sensitive IoT}} & \multirow{2}{*}{\shortstack{ MEC-empowered NTN}}  & \multirow{2}{*}{\shortstack{Oasis-oriented under a \\ cell-free architecture}}& \\
		 &  &  & \\
		\bottomrule[1pt]
	\end{tabular}  
\end{table}  


\subsection{Main Contributions}
\par{
In this paper, the design of an MEC-empowered NTN is studied for wide-area time-sensitive IoT. Particularly, we focus on the design of NTN, which consists of hierarchically integrated satellite and UAVs, while considering the overall communication and computing latency as the metric of latency performance. To satisfy the service requirements of wide-area time-sensitive IoT, the MEC-empowered NTN is designed under a process-oriented framework in a time-division manner \cite{Liu2020}, where a latency minimization problem is formulated using the large-scale CSI. Then, a process-oriented joint resource orchestration scheme is proposed to solve the latency minimization problem. Concretely, the main contributions of this paper are summarized as follows.
\begin{itemize}
	\item A process-oriented framework is presented for the MEC-empowered NTN. This framework can jointly design the communication and MEC systems in a time-division manner for hierarchically integrated satellite and UAVs. After that, an overall communication and computing latency minimization problem is formulated, where the large-scale CSI is used to characterize the complex propagation environment with an affordable cost. 
	\item As the latency minimization problem is a non-convex stochastic optimization problem, we propose an approximation of this problem. The approximated problem can be further decomposed into subproblems according to the properties of the overall communication and computing efficiency function.
	\item To provide solutions to the subproblems, we propose a joint power allocation and data stream scheduling scheme, where block coordinate descent and successive convex approximation techniques are applied. Furthermore, the process-oriented joint resource orchestration scheme is derived in an iterative way.
\end{itemize}
}
\par{
The rest of this paper is organized as follows. We introduce the system model and process-oriented framework in Section~II. In Section~III, we investigate the solution to the overall communication and computing latency minimization problem, where a joint power allocation and data stream scheduling scheme is proposed. Simulation results and discussions are presented in Section~IV, while the conclusions of this paper are drawn in Section~V.
}

\section{System Model}
\begin{figure*}[t]
	\centering
	\includegraphics[width=5.0in]{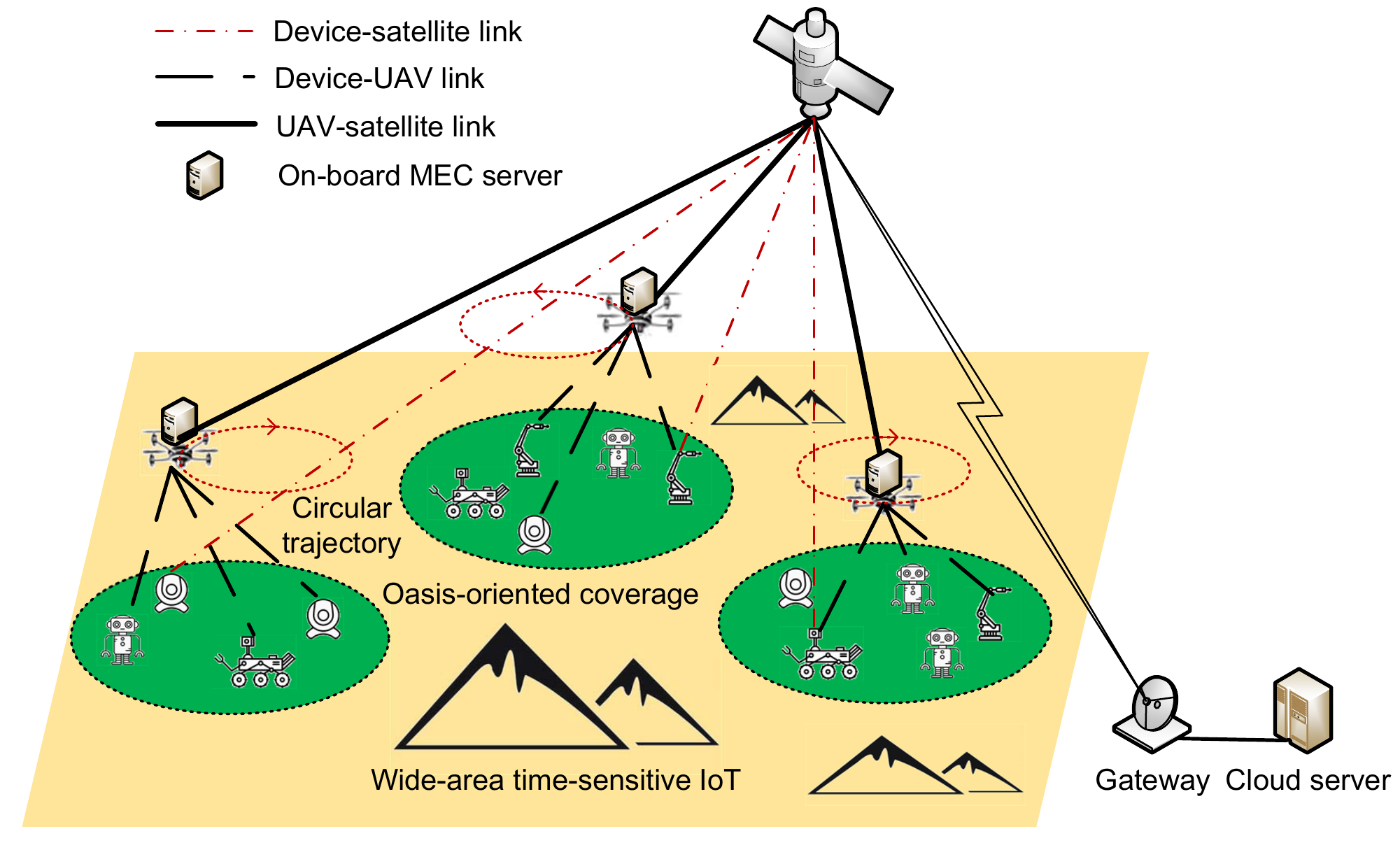}
	\caption{ Illustration of an MEC-empowered hierarchical NTN for wide-area time-sensitive IoT.}
	\label{fig1}
\end{figure*}
\par{
	We consider an MEC-empowered NTN with hierarchically integrated satellite and UAVs as shown in Fig. 1, where there are $U$ single-antenna IoT devices, $K$ UAVs in a swarm where each UAV is equipped with an MEC server and $M$ antennas, and a satellite that can transmit data back to the cloud server via a gateway. We assume that the UAVs are flying around IoT devices following a predetermined circular trajectory, which can save energy with guaranteed stability \cite{Chen2018}. To accommodate the distribution of devices in wide-area time-sensitive IoT, the hierarchical NTN is designed under a cell-free architecture \cite{Liu2020}, where an oasis-oriented coverage pattern can be observed. Based on such coverage pattern, the associations between the devices and UAVs can be predetermined. For the convenience of mathematical analysis, the indicator set of user association is denoted as $\mathbf{z}=\{z_{u, k}\}$, where $z_{u, k} = 1$ means that the $u$-th device is associated with the $k$-th UAV.
}
\begin{figure}[t]
	\centering
	\includegraphics[width=3.0in]{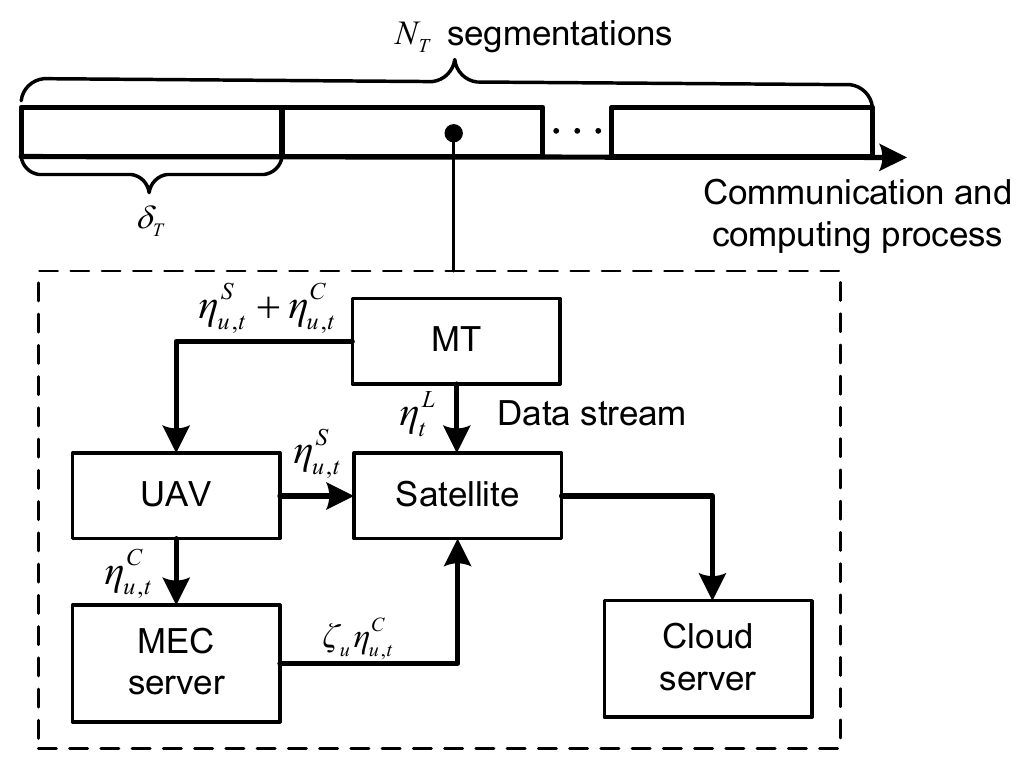}
	\caption{ Diagram of the process-oriented framework in the MEC-empowered NTN.}
	\label{fig2}
\end{figure}
\par{
 In practical systems, the computing ability of each IoT device is usually weak, so the devices must upload data to the satellite or UAVs to accomplish computation-intensive but time-sensitive tasks \cite{Cheng2019}. After the cloud server successfully receives all the data from the devices, the communication and computing process is finished \cite{Pan2020, Wang2020TVT}. The $u$-th device is assumed to have $D_u$ data to upload. To cope with the influence of UAV movement on data transmission, the communication and computing process is designed under a process-oriented framework, which can reduce the complexity of optimizing the whole process \cite{Liu2020}. As illustrated in Fig.~\ref{fig2}, the whole process is divided into $N_T$ segmentations. At the beginning of a segmentation, parameters of the MEC-empowered NTN are updated, and these parameters are assumed to be constant during each segmentation and may vary with each other in different segmentations. Denoting the update interval of system parameters as $\delta_T$, the overall communication and computing latency can be expressed as $T_{total} = N_T \delta_T + \epsilon_a$ , where $\epsilon_a$ denotes the total propagation time of the electromagnetic wave. More specifically, in the $t$-th segmentation, the $u$-th IoT device can send a ratio of $\eta^L_t$ data directly to satellite, a ratio of $\eta_{u,t}^{S}$ data to satellite via device-UAV and UAV-satellite links, and a ratio of $\eta_{u,t}^{C}$ data via device-UAV links into on-board MEC servers for computing. Thus, we have
 \begin{align} 
 & \eta_{t}^L + \eta_{u, t}^S + \eta_{u, t}^C = 1 \ \ \forall u,t \label{eta1}  \\
 & 0 \leq \eta_{t}^L,\eta_{u, t}^S,\eta_{u, t}^C \leq 1 \ \ \forall u,t \label{eta2}
 \end{align} 
 as practical constraints of these ratios. Particularly, after the data are computed by MEC server, the results of computing are transmitted from MEC servers to satellite via UAV-satellite links. For the convenience of mathematical analysis, we assume that the size of the output data is proportional to that of input data for MEC servers \cite{Wang2019iotj, Sharma2018}, where the proportion of the data from the $u$-th device is denoted as $\zeta_u$.
}
\par{
	In each segmentation of the process, the data from IoT devices are transmitted to the satellite or UAVs at first. Under the cell-free architecture, all devices are assumed to share the same frequency band \cite{Liu2020}, where the bandwidth is denoted as $B$. When an IoT device is directly connected with the satellite, we assume that the UL rate between the device and satellite is a constant \cite{Dymond1988}, which is $R^L$. Moreover, when IoT devices are connected with UAVs, they consist of a multi-user multiple-input-multiple-output (MU-MIMO) UL system for data transmission. Therefore, the received symbol of the $u$-th user from the $k$-th UAV in the $t$-th segmentation is formulated as
\begin{equation} \label{signal}
	\mathbf{y}_{u, k, t} =  \mathbf{h}_{u, k, t} x_{u, t} + \sum_{v = 1, v \neq u}^{U} \mathbf{h}_{v, k, t} x_{v, t} + \mathbf{n}_{u, k, t}
\end{equation}
where $x_{u, t}$ is the transmitted symbol, $\mathbf{n}_{u, k, t}$ denotes the additive white Gaussian noise which satisfies $\mathbf{n}_{u, k, t} \sim \mathcal{N}(0, \sigma^2 \mathbf{I}_{M})$, $\mathbf{h}_{u, k, t}$ is the channel vector, which is written as
\begin{equation} \label{channelmodel}
	\mathbf{h}_{u, k, t} = s_{u, k, t} l_{u, k} \mathbf{a}_{u, k}.
\end{equation}
In (\ref{channelmodel}), $s_{u, k, t}$ denotes the fast-varying small-scale parameters with identical distribution, whose phase is uniformly distributed in $[0, 2\pi]$, while amplitude obeys Nakagami-m distribution, where the probability density function is \cite{Chen2018}
\begin{equation} \label{Nakagamipdf}
f_{|s|}(z)=\frac{2m^m}{\Gamma(m)\Omega^m}z^{2m-1} e^{-\frac{mz^2}{\Omega}}.
\end{equation}
In (\ref{Nakagamipdf}), $m$ and $\Omega$ are parameters of Nakagami-m distribution, $\Gamma(m)$ denotes the Gamma function with respect to $m$.
These small-scale parameters are assumed to be independent with each other for different $u$, $k$ and $t$ \cite{Khuwaja2018}. Besides, $l_{u, k}$ is the slowly-varying path loss of UAV channel, which is expressed as follows \cite{Chen2018}:
\begin{equation}
	l_{u, k} = 10^{-\frac{L_{u,k}}{20}}
\end{equation}
\begin{equation}
	L_{u,k} = \frac{A_0}{1 + a e^{-b(\theta_{u, k} - a)}} + B_0
\end{equation}
where $A_0 = \eta_{\textnormal{LOS}} - \eta_{\textnormal{NLOS}}$, $B_0 = 20 \log_{10} (d_{u, k}) + 20\log_{10}(4\pi f/c) + \eta_{\textnormal{NLOS}}$, $f$ is the carrier frequency, $c$ denotes the speed of light, $\eta_{\textnormal{LOS}}$, $\eta_{\textnormal{NLOS}}$, $a$ and $b$ are constants related to propagation environment, $d_{u, k}$ denotes the distance between the device and UAV, $\theta_{u, k} = \frac{180}{\pi} \arcsin(\frac{h_{k}}{d_{u, k}})$ represents the azimuth angle of the device-UAV link, $h_{k}$ is the height of $k$-th UAV. Moreover, $\mathbf{a}_{u, k} \in \mathbb{C}^{M \times 1}$ denotes the array manifold vector of receive antenna array. We assume that uniform linear array (ULA) can be carried on UAVs, so that we have the following expressions \cite{Du2020}:
\begin{equation}
	\mathbf{a}_{u, k} = [1,e^{j\frac{2\pi f d_0}{c} \cos(\theta_{u, k})} ,..., e^{j\frac{2\pi f d_0}{c}(M - 1)\cos(\theta_{u, k})}]^{T}
\end{equation}
where $d_0$ is the distance between adjacent antennas. In this MU-MIMO system, the process of data transmission from devices to UAVs is designed prior to UAV take-off, and this has a much larger time scale than the channel coherence time. Therefore, using pilot symbols, the UL CSI in (\ref{channelmodel})  can be accurately estimated by UAVs within the channel coherence time, but such CSI cannot be perfectly acquired prior to UAV take-off with respect to the large time scale of the whole process. As a result, perfect CSI is difficult to use for designing the data transmission process. Specifically, we regard the position-related parameters, i.e., $l_{u, k}$ and $\mathbf{a}_{u, k}$ as slowly varying large-scale channel parameters, which can be perfectly acquired using radio maps in practical systems \cite{Li2020}. These parameters are assumed to be constant during the whole process. In contrast, $s_{u, k,t}$ varies rapidly due to the movement of UAVs, and only its distribution is known. Under these assumptions, the efficiency of data transmission in each segmentation of the process can be evaluated by the ergodic rate \cite{Liu2020}. In addition, it is reasonable to assume that minimum mean square error (MMSE) detection is used at the receiver \cite{Pan2020}, where the detection vector for the $u$-th device at the $k$-th UAV in the $t$-th segmentation is denoted as $\mathbf{w}_{u, k, t}$ \cite{Ammari2017}. As a result, the UL ergodic rate of the $u$-th device at the $k$-th UAV in the $t$-th segmentation can be formulated as follows \cite{Cao2016}:
\begin{align} \label{RU}
\nonumber	& R^{U}_{u, k, t}(\mathbf{P}) = (1 - \gamma_{\textnormal{UL}})B \\
&\mathbf{E} \left\{ \log_{2}\left( 1 + \frac{p_{u, t} |\mathbf{w}^H_{u, k, t} \mathbf{h}_{u, k, t}|^2}{\sum_{v = 1, v \neq u}^{U} p_{v, t} |\mathbf{w}^H_{u, k, t} \mathbf{h}_{v, k, t}|^2 + ||\mathbf{w}_{u, k, t}||^2 \sigma^2} \right) \right\}
\end{align}
where $\gamma_{\textnormal{UL}}$ denotes the fraction of transmitted signals that are used as reference signals, which which means that $R^{U}_{u, k, t}(\mathbf{P})$ can be regarded as the achievable net rate \cite{Rost2012}, $\mathbf{P} = (p_{u, t}) \in \mathbb{R}^{U \times N_{T}}$ denotes power matrix, $p_{u, t} = \mathbf{E}\{x^{H}_{u,t}x_{u,t}\}$ is signal power.
}
\par{After the UAVs receive the data streams from IoT devices, these streams are further scheduled for communication and computing. To guarantee the stability, the system is assumed to work in a non-blocking mode \cite{Wang2019iotj},where any packet of data in the data stream can be transmitted from devices to the cloud server without any waiting time. Therefore, the constraints of the data streams are derived as follows \cite{Wang2019iotj}:
\begin{align}
	& \sum_{u = 1}^U \eta_{u,t}^C z_{u, k} R^U_{u,k,t}(\mathbf{P}) \leq R_k^C \ \ \forall k \label{constraintRC} \\
	& \sum_{u=1}^U (\eta_{u,t}^S + \zeta_u \eta_{u,t}^C) z_{u, k} R^U_{u, k, t}(\mathbf{P}) \leq R_k^S \ \ \forall k \label{constraintRS}
\end{align}
 where $R_k^C$ denotes the average throughput of MEC server, $R_k^S$ is the data rate of UAV-satellite link, both of them are parameters of the $k$-th UAV. Furthermore, for the $u$-th device in the $t$-th segmentation, the average overall communication and computing efficiency is expressed as follows:
 \begin{itemize}
 	\item If only the satellite is used for communication, we have
 	\begin{equation} \label{Ra1}
 		R^a_{u, t} = R^L
 	\end{equation}
 	\item If the MEC server on UAV is not used for computing, we have
 	\begin{align} \label{Ra2}
 	\nonumber R^a_{u, t} =  &\left( \frac{\eta^L_{t}}{R^L} + \frac{\eta^{S}_{u, t}}{\sum_{k = 1}^{K} z_{u, k} R^U_{u,k,t}(\mathbf{P})} \right.  \\
 	& \left. + \frac{\eta^{S}_{u, t}}{\sum_{k = 1}^{K} z_{u, k} R_k^S}  \right)^{-1}
 	\end{align}
 	\item If the MEC server on UAV is used for computing, we have
 	\begin{align} \label{Ra3}
 	\nonumber	R^a_{u, t} =  & \left( \frac{\eta^L_{t}}{R^L} + \frac{\eta^{S}_{u, t} + \eta^{C}_{u, t}}{\sum_{k = 1}^{K} z_{u, k} R^U_{u,k,t}(\mathbf{P})} \right.  \\
 		& \left. + \frac{\eta^{S}_{u, t} + \zeta_u \eta^{C}_{u, t}}{\sum_{k = 1}^{K} z_{u, k} R_k^S} + \frac{\eta^{C}_{u, t}}{\sum_{k = 1}^{K} z_{u, k} R_k^C} \right)^{-1}
 	\end{align}
 \end{itemize}
where $\epsilon_0$ is the transmission time of a data packet. In (\ref{Ra1})--(\ref{Ra3}), $R^a_{u, t}$ has different values in different cases. The reason for this phenomenon is that in practical systems, data are transmitted from devices to satellite and UAVs in sequential packets. To keep the system stable, update interval should be longer than the packet transmission time. As indicated by Fig.~\ref{fig2}, it requires at least $2\epsilon_0$ packet transmission time to send data to satellite via device-UAV links and UAV-satellite links, while at least $3 \epsilon_0$ packet transmission time is consumed to send data to satellite via device-UAV links, MEC servers and UAV-satellite links. Thus, UAVs and MEC servers can only be used when the update interval is sufficiently large, which makes $R^a_{u, t}$ become a piecewise function.
}
\par{On the basis of (\ref{eta1})--(\ref{Ra3}), the overall communication and computing latency minimization problem under a process-oriented framework can be formulated as 
\begin{subequations}\label{pro0.1}
	\begin{align}
	\min_{\mathbf{P}, \bm{\eta}, \delta_{T}} & \ \delta_{T} \label{pro0.1a} \\
	s.t. \ & \sum_{t = 1}^{N_{T}} R^{a}_{u, t} \delta_T \geq D_u \ \forall u \label{pro0.1aa} \\
	& \sum_{u = 1}^U \eta_{u, t}^C z_{u, k} R^U_{u,k,t}(\mathbf{P}) \leq R_k^C \ \ \forall k,t \label{pro0.1b} \\
	& \sum_{u=1}^U (\eta_{u, t}^S + \zeta_u \eta_{u, t}^C) z_{u, k} R^U_{u, k,t}(\mathbf{P}) \leq R_k^S \ \ \forall k,t \label{pro0.1c}\\
	& 0 \leq p_{u, t} \leq P_{max} \ \ \forall u, t \label{pro0.1d}\\
	& \eta_{t}^L + \eta_{u, t}^S + \eta_{u, t}^C = 1 \ \ \forall u,t \label{pro0.1e} \\
	& 0 \leq \eta_{t}^L,\eta_{u, t}^S,\eta_{u, t}^C \leq 1 \ \ \forall u,t \label{pro0.1f}
	\end{align}
\end{subequations}
where $N_T$ is predetermined and $\delta_T$ is optimized to find the minimum overall latency, (\ref{pro0.1aa}) can guarantee that the service requirements of all IoT devices are satisfied, (\ref{pro0.1b}) and (\ref{pro0.1c}) are rate constraints of data streams, (\ref{pro0.1d}) denotes the power constraints of the devices where $P_{max}$ is the maximum transmit power, (\ref{pro0.1e}) represents the practical constraints of data stream scheduling and (\ref{pro0.1f}) shows the range of variables. 
}

\section{ Joint Power Allocation and Data Stream Scheduling under the Process-Oriented Framework}
\subsection{Problem Decomposition}
\par{
Observing (\ref{pro0.1aa})--(\ref{pro0.1c}), we can find that (\ref{pro0.1}) is a non-convex stochastic optimization problem, which is difficult to solve directly using existing tools. To further simplify (\ref{pro0.1}), an approximation of $R^U_{u,k,t}(\mathbf{P})$ is given as follows \cite{Khoshnevis2013,Cao2016}.
\begin{align} \label{approrate}
\nonumber &\hat{R}^{U}_{u, k, t}(\mathbf{P}) = (1-\gamma_{\textnormal{UL}})B \\
&\log_{2}\left( 1 + \frac{p_{u, t} \theta_{u, u, k}}{\sum_{v = 1, v \neq u}^{U} p_{v, t} \theta_{u, v, k} + \sigma^2} \right)
\end{align}
where 
\begin{equation} \label{thetauvk}
\theta_{u, v, k} = \mathbf{E} \left\{ \frac{|\mathbf{w}^H_{u, k, t} \mathbf{h}_{v, k, t}|^2}{||\mathbf{w}_{u, k, t}||^2} \right\}.
\end{equation}
According to (\ref{approrate}) and (\ref{thetauvk}), $\bm{\theta} = \{\theta_{u, v, k} \ \forall u,v,k\}$ is calculated prior to resource orchestration, which can be regarded as deterministic parameters, indicating that the expectation operator in (\ref{approrate}) could be eliminated. With randomly generated channel vectors and transmit power, it is demonstrated in Fig.~\ref{fig3} that the proposed approximation is accurate, while a justification of such approximation is given in Appendix A.
}
\begin{figure}[t]
	\centering
	\includegraphics[width=3.8in]{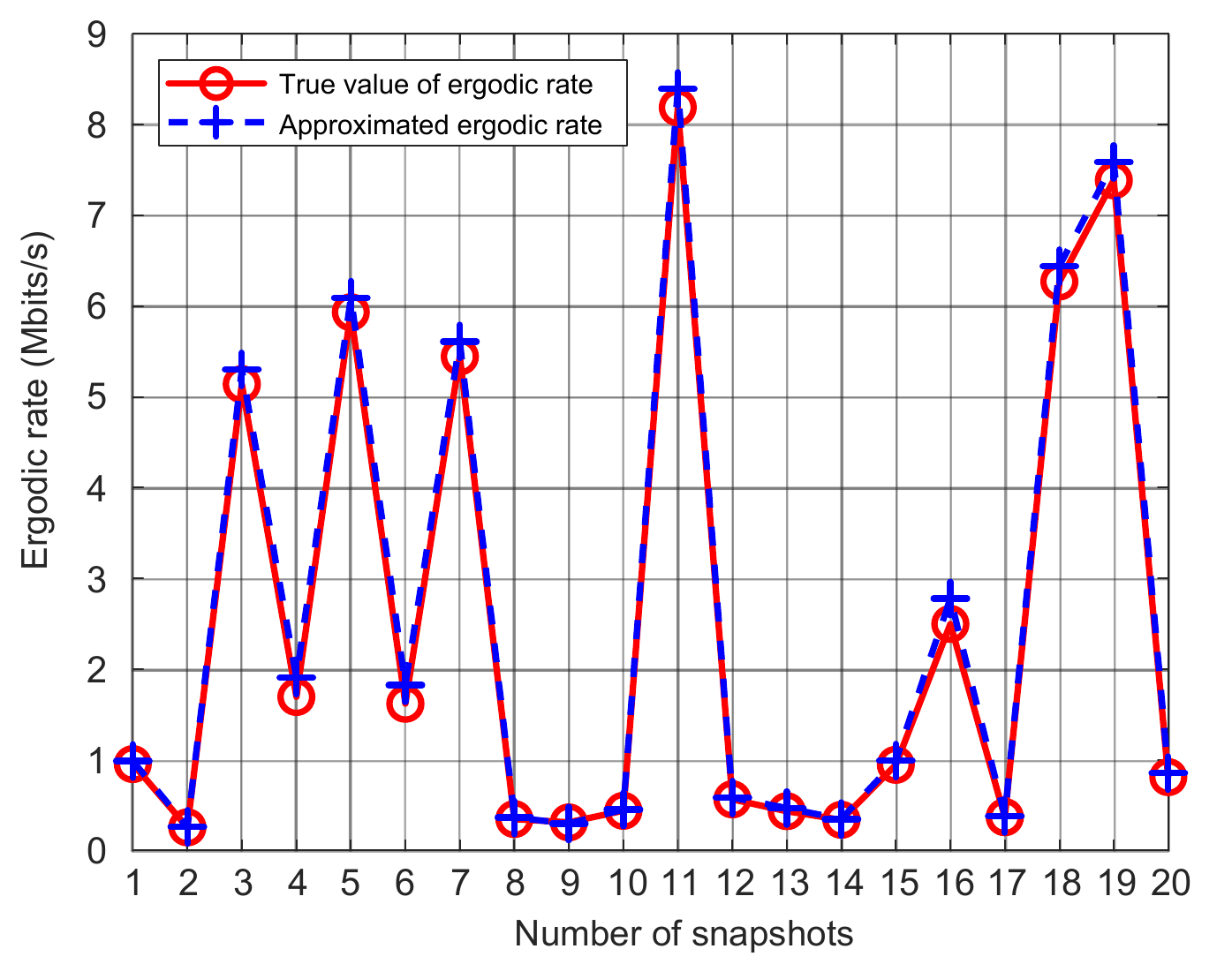}
	\caption{Numerical evaluations of the accuracy of approximated ergodic rate.}
	\label{fig3}
\end{figure}
\par{
Based on (\ref{approrate}) and (\ref{thetauvk}), $R^U_{u,k,t}(\mathbf{P})$ in (\ref{pro0.1aa})--(\ref{pro0.1c}) can be replaced by $\hat{R}^U_{u,k,t}(\mathbf{P})$. Then, (\ref{pro0.1}) is further decomposed into three subproblems regarding the segmentation of $R^{a}_{u, t}$, which are illustrated as follows.
	\begin{subequations}\label{pro1}
		\begin{align}
		\min_{\mathbf{P}, \bm{\eta}, \delta_T } & \ \delta_T \label{pro1a} \\
		s.t. \ & N_{T} \delta_{T} R^L \geq D_u \ \forall u \label{pro1aa} \\
		& \sum_{u = 1}^U \eta_{u, t}^{C} z_{u, k} 
		\hat{R}^U_{u,k,t}(\mathbf{P}) \leq R^C_k \ \ \forall k,t \label{pro1b} \\
		& \sum_{u=1}^U (\eta_{u, t}^{S} + \zeta_u \eta_{u, t}^{C}) z_{u, k} \hat{R}^U_{u, k, t}(\mathbf{P}) \leq R^S_k \ \ \forall k,t \label{pro1c}\\
		& 0 \leq p_{u, t} \leq P_{max} \ \ \forall u,t \label{pro1d} \\
		& \eta_{t}^L + \eta_{u, t}^S + \eta_{u, t}^C = 1 \ \ \forall u,t \label{pro1e} \\
		& 0 \leq \eta_{t}^L,\eta_{u, t}^S,\eta_{u, t}^C \leq 1 \ \ \forall u,t \label{pro1f} \\
		&  \delta_T \geq \epsilon_0 \label{pro1g}
		\end{align}
	\end{subequations}
	\begin{subequations}\label{pro2}
		\begin{align}
		\min_{\mathbf{P}, \bm{\eta}, \delta_T} & \ \delta_T \label{pro2a} \\
		\nonumber s.t. \ & \sum_{t=1}^{N_{T}} \delta_{T} \left( \frac{\eta^L_{t}}{R^L} + \frac{\eta^{S}_{u, t} + \eta^{C}_{u, t}}{\sum_{k = 1}^{K} z_{u, k} \hat{R}^U_{u,k,t}(\mathbf{P})} \right.  \\
		\nonumber & \left. + \frac{\eta^{S}_{u, t} + \zeta_u \eta^{C}_{u, t}}{\sum_{k = 1}^{K} z_{u, k} R_k^S} + \frac{\eta^{C}_{u, t}}{\sum_{k = 1}^{K} z_{u, k} R_k^C} \right)^{-1} \\
		& \geq D_u \ \forall u \label{pro2aa} \\
		& \sum_{u = 1}^U \eta_{u, t}^{C} z_{u, k} 
		\hat{R}^U_{u,k,t}(\mathbf{P}) \leq R^C_k \ \ \forall k,t \label{pro2b} \\
		& \sum_{u=1}^U (\eta_{u, t}^{S} + \zeta_u \eta_{u, t}^{C}) z_{u, k} \hat{R}^U_{u, k, t}(\mathbf{P}) \leq R^S_k \ \ \forall k,t \label{pro2c}\\
		& 0 \leq p_{u, t} \leq P_{max} \ \ \forall u,t \label{pro2d} \\
		& \eta_{t}^L + \eta_{u, t}^S + \eta_{u, t}^C = 1 \ \ \forall u,t \label{pro2e} \\
		& 0 \leq \eta_{t}^L,\eta_{u, t}^S,\eta_{u, t}^C \leq 1 \ \ \forall u,t \label{pro2f} \\
		& \delta_T \geq 3 \epsilon_0 \label{pro2g}
		\end{align}
	\end{subequations}
	\begin{subequations}\label{pro3}
		\begin{align}
		\min_{\mathbf{P}, \bm{\eta}, \delta_T} & \ \delta_T \label{pro3a} \\
		\nonumber s.t. \ & \sum_{t=1}^{N_{T}} \delta_{T} \left( \frac{\eta^L_{t}}{R^L} + \frac{\eta^{S}_{u, t}}{\sum_{k = 1}^{K} z_{u, k} \hat{R}^U_{u,k,t}(\mathbf{P})} \right.  \\
		& \left. + \frac{\eta^{S}_{u, t}}{\sum_{k = 1}^{K} z_{u, k} R_k^S}  \right)^{-1} \geq D_u \ \forall u \label{pro3aa} \\
		& \sum_{u = 1}^U \eta_{u, t}^{C} z_{u, k} 
		\hat{R}^U_{u,k,t}(\mathbf{P}) \leq R^C_k \ \ \forall k,t \label{pro3b} \\
		& \sum_{u=1}^U (\eta_{u, t}^{S} + \zeta_u \eta_{u, t}^{C}) z_{u, k} \hat{R}^U_{u, k, t}(\mathbf{P}) \leq R^S_k \ \ \forall k,t \label{pro3c}\\
		& 0 \leq p_{u, t} \leq P_{max} \ \ \forall u,t \label{pro3d} \\
		& \eta_{t}^L + \eta_{u, t}^S + \eta_{u, t}^C = 1 \ \ \forall u,t \label{pro3e} \\
		& 0 \leq \eta_{t}^L,\eta_{u, t}^S,\eta_{u, t}^C \leq 1 \ \ \forall u,t \label{pro3f} \\
		& \delta_T \geq 2 \epsilon_0. \label{pro3g}
		\end{align}
	\end{subequations}
It is not difficult to certify that the subproblems in (\ref{pro1})--(\ref{pro3}) are independent with each other. For the simplification of notations, if we have $(\mathbf{P}, \bm{\eta})$ as the solution to (\ref{pro1}), (\ref{pro2}) or (\ref{pro3}), the corresponding objective function is expressed as $\delta_T(\mathbf{P}, \bm{\eta})$.  Then, we discuss the solutions to these subproblems respectively.
}

\subsection{The Solution to (\ref{pro1})}
\par{
According to the constraints in (\ref{pro1}), we can find that (\ref{pro1b})--(\ref{pro1f}) have no influence on the objective function in (\ref{pro1a}). Therefore, (\ref{pro1}) can be equivalently transformed to 
	\begin{subequations}\label{pro1.1}
	\begin{align}
	\min_{\mathbf{P}, \bm{\eta}, \delta_T } & \ \delta_T \label{pro1.1a} \\
	s.t. \ & N_{T} \delta_{T} R^L \geq D_u \ \forall u \label{pro1.1b} \\
	& \delta_T \geq \epsilon_0 \label{pro1.1c}
	\end{align}
\end{subequations}
whose solution can be directly given by
\begin{equation} \label{pro1eqn1}
	 \eta^L_t = 1, \eta^S_{u, t} = \eta^C_{u, t} = 0, p_{u, t} = P_{max}
\end{equation}
and the minimum update interval of process becomes
\begin{equation} \label{pro1eqn2}
	\delta_T = \max \left\{ \epsilon_0, \frac{D_u}{N_T R^{L}} \ \forall u \right\}.
\end{equation}

}
\begin{remark}
	\textnormal{ The solution to (\ref{pro1}) provides the joint resource orchestration scheme when we only use the satellite to transmit data. Intuitively, we point out that the minimum communication and computing latency is achieved by this strategy when $D_u$ is sufficiently small. The reason is that the communication and computing latency may be less than $2 \epsilon_0$ if we only use the satellite for data transmission, but the latency is at least $2 \epsilon_0$ if UAVs are used, as shown in (\ref{pro1g}), (\ref{pro2g}) and (\ref{pro3g}). Such intuition can be further certified by simulation results.
}
\end{remark}

\subsection{The Solution to (\ref{pro2})}
\par{
	Due to the coupling of $\mathbf{P}$ and $\bm{\eta}$ in (\ref{pro2aa})--(\ref{pro2c}), (\ref{pro2}) is non-convex and hard to be solved directly. To handle this problem, we decompose (\ref{pro2}) into two subproblems following the block coordinate descent technique \cite{Liu2020}, which are formulated as
		\begin{subequations}\label{pro2.1}
			\begin{align}
			\min_{\mathbf{P}^{i}, \delta_T } & \ \delta_T \label{pro2.1a} \\
			\nonumber s.t. \ & \sum_{t=1}^{N_{T}} \delta_{T} \left( \frac{\eta^{L, i - 1}_{t}}{R^L} + \frac{\eta^{S, i - 1}_{u, t} + \eta^{C, i - 1}_{u, t}}{\sum_{k = 1}^{K} z_{u, k} \hat{R}^U_{u,k,t}(\mathbf{P}^i)} \right.  \\
			\nonumber & \left. + \frac{\eta^{S, i - 1}_{u, t} + \zeta_u \eta^{C, i - 1}_{u, t}}{\sum_{k = 1}^{K} z_{u, k} R_k^S} + \frac{\eta^{C, i - 1}_{u, t}}{\sum_{k = 1}^{K} z_{u, k} R_k^C} \right)^{-1} \\
			& \geq D_u \ \forall u \label{pro2.1b} \\
			& \sum_{u = 1}^U \eta_{u, t}^{C, i - 1} z_{u, k} 
			\hat{R}^U_{u,k,t}(\mathbf{P}^{i}) \leq R^C_k \ \ \forall k,t \label{pro2.1c} \\
			& \sum_{u=1}^U (\eta_{u, t}^{S, i - 1} + \zeta_u \eta_{u, t}^{C, i - 1}) z_{u, k} \hat{R}^U_{u, k, t}(\mathbf{P}^{i}) \leq R^S_k \ \ \forall k,t \label{pro2.1d}\\
			& 0 \leq p^{i}_{u, t} \leq P_{max} \ \ \forall u,t \label{pro2.1e} \\
			& \delta_T \geq 3 \epsilon_0 \label{pro2.1f}
			\end{align}
		\end{subequations}
		\begin{subequations}\label{pro2.2}
			\begin{align}
			\min_{\bm{\eta}^{i}, \delta_T} & \ \delta_T \label{pro2.2a} \\
			\nonumber s.t. \ & \sum_{t=1}^{N_{T}} \delta_{T} \left( \frac{\eta^{L, i}_{t}}{R^L} + \frac{\eta^{S, i}_{u, t} + \eta^{C, i}_{u, t}}{\sum_{k = 1}^{K} z_{u, k} \hat{R}^U_{u,k,t}(\mathbf{P}^{i})} \right.  \\
			\nonumber & \left. + \frac{\eta^{S, i}_{u, t} + \zeta_u \eta^{C, i}_{u, t}}{\sum_{k = 1}^{K} z_{u, k} R_k^S} + \frac{\eta^{C, i}_{u, t}}{\sum_{k = 1}^{K} z_{u, k} R_k^C} \right)^{-1} \\
			& \geq D_u \ \forall u \label{pro2.2aa} \\
			& \sum_{u = 1}^U \eta_{u, t}^{C, i} z_{u, k} 
			\hat{R}^U_{u,k,t}(\mathbf{P}^{i}) \leq R^C_k \ \ \forall k,t \label{pro2.2b} \\
			& \sum_{u=1}^U (\eta_{u, t}^{S, i} + \zeta_u \eta_{u, t}^{C, i}) z_{u, k} \hat{R}^U_{u, k, t}(\mathbf{P}^{i}) \leq R^S_k \ \ \forall k,t \label{pro2.2c}\\
			& \eta_{t}^{L, i} + \eta_{u, t}^{S, i} + \eta_{u, t}^{C, i} = 1 \ \ \forall u,t \label{pro2.2d} \\
			& 0 \leq \eta_{t}^{L, i},\eta_{u, t}^{S, i},\eta_{u, t}^{C, i} \leq 1 \ \ \forall u,t \label{pro2.2e} \\
			& \delta_T \geq 3 \epsilon_0 \label{pro2.2f}
			\end{align}
		\end{subequations}
 where $i$ is denoted as iteration index, (\ref{pro2.1}) is the power allocation subproblem and (\ref{pro2.2}) denotes the data stream scheduling subproblem. After these, we discuss the solutions to (\ref{pro2.1}) and (\ref{pro2.2}) respectively.
}
\subsubsection{The solution to (\ref{pro2.1})}
\par{
	At first, the properties of $\hat{R}^U_{u, k, t}(\mathbf{P})$ should be clarified in order to give the solution to (\ref{pro2.1}).
	\begin{theorem}
		$\hat{R}^U_{u, k, t}(\mathbf{P})$ is concave with respect to $p_{u, t}$ and convex with respect to $\{p_{v, t} \ \ \forall v \neq u\}$. Accordingly, at any given point $\mathbf{P}^0$, we have
		\begin{equation} \label{th2.1}
		\hat{R}^U_{u, k, t}(\mathbf{P}) \geq (1-\gamma_{\textnormal{UL}}) B  \bar{R}^U_{u, k}(p_{u, t}, I_{u, k}(\bar{\mathbf{p}}_{u, t}) | p^0_{u, t}, I_{u, k}(\bar{\mathbf{p}}^0_{u, t}))
		\end{equation}
		\begin{equation} \label{th2.2}
		\hat{R}^U_{u, k, t}(\mathbf{P}) \leq (1-\gamma_{\textnormal{UL}}) B  \tilde{R}^U_{u, k}(p_{u, t}, I_{u, k}(\bar{\mathbf{p}}_{u, t}) | p^0_{u, t}, I_{u, k}(\bar{\mathbf{p}}^0_{u, t}))
		\end{equation}
		where
		\begin{equation}
		I_{u, k}(\bar{\mathbf{p}}_{u, t}) = \sum_{v = 1,v \neq u}^{U} p_{v, t} \theta_{u, v, k} + \sigma^2, \bar{\mathbf{p}}_{u, t} = \{p_{v, t}, v \neq u\}
		\end{equation}
		\begin{align} \label{Rbar}
		\nonumber	&\bar{R}^U_{u, k}(p_{u, t}, I_{u, k}(\bar{\mathbf{p}}_{u, t}) | p^0_{u, t}, I_{u, k}(\bar{\mathbf{p}}^0_{u, t})) \\
		\nonumber	&= \textnormal{log}_2(p_{u, t} \theta_{u,u,k} + I_{u, k}(\bar{\mathbf{p}}_{u, t}) + \sigma^2) - \textnormal{log}_2( I_{u, k}(\bar{\mathbf{p}}^0_{u, t}) + \sigma^2)\\
		& - \frac{1}{\textnormal{ln}2 (I_{u, k}(\bar{\mathbf{p}}^0_{u, t}) + \sigma^2)}(I_{u, k}(\bar{\mathbf{p}}_{u, t})- I_{u, k}(\bar{\mathbf{p}}^0_{u, t}))
		\end{align}
		\begin{align} \label{Rsim}
		\nonumber &\tilde{R}^U_{u, k}(p_{u, t}, I_{u, k}(\bar{\mathbf{p}}_{u, t}) | p^0_{u, t}, I_{u, k}(\bar{\mathbf{p}}^0_{u, t})) \\
		\nonumber & = \textnormal{log}_2(p^0_{u, t} \theta_{u,u,k} + I_{u, k}(\bar{\mathbf{p}}_{u, t}^0) + \sigma^2) \\ 
		\nonumber & +\frac{1}{\textnormal{ln}2 (p^0_{u, t} \theta_{u,u,k} + I_{u, k}(\bar{\mathbf{p}}_{u, t}^0) + \sigma^2)} (p_{u, t} \theta_{u,u,k} - p^0_{u, t} \theta_{u,u,k} \\
		& + I_{u, k}(\bar{\mathbf{p}}_{u, t}) -   I_{u, k}(\bar{\mathbf{p}}_{u, t}^0)) - \textnormal{log}_2 (I_{u, k}(\bar{\mathbf{p}}_{u, t}) + \sigma^2).
		\end{align}
	\end{theorem}
	\begin{IEEEproof}
		See Appendix B.
	\end{IEEEproof}
	According to Theorem 1, although (\ref{pro2.1}) is non-convex, it can be solved iteratively using successive convex approximation technique \cite{Sun2017}. Denoting the iteration index as $j$, the problem in (\ref{pro2.1}) is recast to
		\begin{subequations}\label{pro2.1.1}
			\begin{align}
			\min_{\mathbf{P}^{i, j}, \delta_T } & \ \delta_T \label{pro2.1.1a} \\
			\nonumber s.t. \ & \frac{D_u}{\delta_T} - \sum_{t = 1}^{N_{T}} \\
			& \frac{\bar{R}^U_{u, t}(\mathbf{P}^{i, j} | \mathbf{P}^{i, j - 1})}{C_{u, t}(\bm{\eta}^{i - 1})  \bar{R}^U_{u, t}(\mathbf{P}^{i, j} | \mathbf{P}^{i, j - 1}) + 1 - \eta^{L, i - 1}_t} \leq 0 \ \forall u \label{pro2.1.1b} \\
			\nonumber & \sum_{u = 1}^U \eta_{u, t}^{C, i - 1}
			z_{u, k}\tilde{R}^U_{u, k}(p^{i, j}_{u, t}, I_{u, k}(\bar{\mathbf{p}}^{i, j}_{u, t}) \\
			& | p^{i, j - 1}_{u, t}, I_{u, k}(\bar{\mathbf{p}}^{i, j - 1}_{u, t})) \leq \frac{R^C_k}{(1-\gamma_{\textnormal{UL}}) B} \ \ \forall k,t \label{pro2.1.1c} \\
			\nonumber & \sum_{u=1}^U (\eta_{u, t}^{S, i - 1} + \zeta_u \eta_{u, t}^{C, i - 1}) z_{u, k}\tilde{R}^U_{u, k}(p^{i, j}_{u, t}, I_{u, k}(\bar{\mathbf{p}}^{i, j}_{u, t}) \\
			& | p^{i, j - 1}_{u, t}, I_{u, k}(\bar{\mathbf{p}}^{i, j - 1}_{u, t})) \leq \frac{R^S_k}{(1-\gamma_{\textnormal{UL}}) B} \ \ \forall k,t \label{pro2.1.1d}\\
			& 0 \leq p^{i}_{u, t} \leq P_{max} \ \ \forall u,t \label{pro2.1.1e} \\
			& \delta_T \geq 3 \epsilon_0 \label{pro2.1.1f}
			\end{align}
		\end{subequations}
		where
		\begin{align}
		\nonumber & C_{u, t}(\bm{\eta}^{i - 1}) = \\
		&\frac{\eta^{L, i - 1}_t D_u}{R^L} + \frac{(\eta^{S, i - 1}_{u, t} + \zeta_u \eta^{C, i - 1}_{u, t}) D_u}{\sum_{k = 1}^K z_{u, k} R_k^S} + \frac{\eta^{C, i - 1}_{u, t} D_u}{\sum_{k = 1}^K z_{u, k} R_k^C}
		\end{align}
		\begin{align}
		\nonumber &\bar{R}^U_{u, t}(\mathbf{P}^{i, j} | \mathbf{P}^{i, j - 1}) = (1-\gamma_{\textnormal{UL}}) B\\
		&\sum_{k = 1}^{K} z_{u, k}\bar{R}^U_{u, k}(p^{i, j}_{u, t}, I_{u, k}(\bar{\mathbf{p}}^{i, j}_{u, t}) | p^{i, j - 1}_{u, t}, I_{u, k}(\bar{\mathbf{p}}^{i, j - 1}_{u, t})).
		\end{align}
	Then, the solution to (\ref{pro2.1.1}) is given according to the following theorem.
	\begin{theorem}
		The problem in (\ref{pro2.1.1}) is convex, whose optimal solution is a feasible solution to (\ref{pro2.1}).
	\end{theorem}
	\begin{IEEEproof}
		See Appendix C.
	\end{IEEEproof}
}
\par{
	Theorem 2 shows that (\ref{pro2.1.1}) can be solved using conventional convex optimization tools \cite{Boyd2004}, which also indicates that the solution to (\ref{pro2.1}) can be derived by using the solution to (\ref{pro2.1.1}) in an iterative way. Detailed steps of this method are recorded in Algorithm 1.
}

\addtolength{\topmargin}{0.05in}
\begin{algorithm}[t]
	\caption{Power allocation algorithm to solve (\ref{pro2.1})}
	\hspace*{0.02in} {\bf Input:} $K$, $M$, $U$, $R^L$, $\{R_k^S, R_k^C \ \forall k\}$, $\bm{\theta}$, $\mathbf{z}$, $\{D_u \ \forall u\}$, $N_T$, $\epsilon_0$, $P_{max}$, $\gamma_{\textnormal{UL}}$, $B$, $\bm{\eta}^{i - 1}$, $\mathbf{P}^{i - 1}$.
	\begin{algorithmic}[1]
		\State Initialization:  $\mathbf{P}^{i, 0} = \mathbf{P}^{i - 1}$, $\epsilon = 1 \times 10^{-2}$, $j = 1$.
		\State Solve (\ref{pro2.1.1}), denoting the optimal solution as $(\delta^{*}_T, \mathbf{P}^*)$, set $\mathbf{P}^{i, j} =\mathbf{P}^*$, $\delta_T(\mathbf{P}^{i, j}, \bm{\eta}^{i - 1}) = \delta^{*}_T$;
		\While{$|1 - \frac{\delta_T(\mathbf{P}^{i, j - 1}, \bm{\eta}^{i - 1})}{\delta_T(\mathbf{P}^{i, j}, \bm{\eta}^{i - 1})}| > \epsilon$}
			\State $j = j + 1$;
			\State Solve (\ref{pro2.1.1}), denoting the optimal solution as $(\delta^{*}_T, \mathbf{P}^*)$, set $\mathbf{P}^{i, j} =\mathbf{P}^*$, $\delta_T(\mathbf{P}^{i, j}, \bm{\eta}^{i - 1}) = \delta^{*}_T$;
		\EndWhile
	\end{algorithmic}
	\hspace*{0.02in} {\bf Output:}
	$\mathbf{P}^{i, j}, \delta_T(\mathbf{P}^{i, j}, \bm{\eta}^{i - 1})$. \\
\end{algorithm}

\subsubsection{The solution to (\ref{pro2.2})}
\par{ 
	The problem in (\ref{pro2.2}) is non-convex, because (\ref{pro2.2aa}) is concave with respect to $\mathbf{\eta}^{i}$. Further by using successive convex approximation technique \cite{Sun2017}, denoting the iteration index as $j$, (\ref{pro2.2}) is reformulated as follows.
		\begin{subequations}\label{pro2.2.1}
		\begin{align}
		\min_{\bm{\eta}^{i, j}, \delta_T} & \ \delta_T \label{pro2.2.1a} \\
		\nonumber s.t. \ & \frac{D_u}{\delta_{T}} \\
		\nonumber	& - \sum_{t=1}^{N_{T}}  \left\{ 
		\frac{1}{C_{u, t}(\bm{\eta}^{i, j - 1}) + \frac{1 - \eta_t^{L, i, j - 1}}{\sum_{k = 1}^{K} z_{u, k} \hat{R}^U_{u,k,t}(\mathbf{P}^{i})}}
		\right.  \\
		\nonumber & \left. - \frac{C_{u, t}(\bm{\eta}^{i, j}) - C_{u, t}(\bm{\eta}^{i, j - 1}) - \frac{\eta_t^{L, i, j} - \eta_t^{L, i, j - 1}}{\sum_{k = 1}^{K} z_{u, k} \hat{R}^U_{u,k,t}(\mathbf{P}^{i})} }{\left[C_{u, t}(\bm{\eta}^{i, j - 1}) + \frac{1 - \eta_t^{L, i, j - 1}}{\sum_{k = 1}^{K} z_{u, k} \hat{R}^U_{u,k,t}(\mathbf{P}^{i})}\right]^2} \right\} \\
		& \leq 0 \ \forall u \label{pro2.2.1aa} \\
		& \sum_{u = 1}^U \eta_{u, t}^{C, i, j} z_{u, k} 
		\hat{R}^U_{u,k,t}(\mathbf{P}^{i}) \leq R^C_k \ \ \forall k,t \label{pro2.2.1b} \\
		& \sum_{u=1}^U (\eta_{u, t}^{S, i, j} + \zeta_u \eta_{u, t}^{C, i, j}) z_{u, k} \hat{R}^U_{u, k, t}(\mathbf{P}^{i}) \leq R^S_k \ \ \forall k,t \label{pro2.2.1c}\\
		& \eta_{t}^{L, i, j} + \eta_{u, t}^{S, i, j} + \eta_{u, t}^{C, i, j} = 1 \ \ \forall u,t \label{pro2.2.1d} \\
		& 0 \leq \eta_{t}^{L, i, j},\eta_{u, t}^{S, i, j},\eta_{u, t}^{C, i, j} \leq 1 \ \ \forall u,t \label{pro2.2.1e} \\
		& \delta_T \geq 3 \epsilon_0. \label{pro2.2.1f}
		\end{align}
	\end{subequations}
	We can find that (\ref{pro2.2.1}) is a convex optimization problem with respect to $\delta_T$ and $\bm{\eta}^{i, j}$, which can be solved using conventional convex optimization tools \cite{Boyd2004}. Besides, the relationship between (\ref{pro2.2}) and (\ref{pro2.2.1}) is illustrated as follows.
	\begin{property}
		The optimal solution to (\ref{pro2.2.1}) is a feasible solution to (\ref{pro2.2}).
	\end{property}
	\begin{IEEEproof}
		See Appendix D.
	\end{IEEEproof}
	 On the basis of Property 1, the solution to (\ref{pro2.2}) can be iteratively derived based on the solution to (\ref{pro2.2.1}), which is recorded in Algorithm 2.
}
\addtolength{\topmargin}{0.05in}
\begin{algorithm}[t]
	\caption{Data stream scheduling algorithm to solve (\ref{pro2.2})}
\hspace*{0.02in} {\bf Input:} $K$, $M$, $U$, $R^L$, $\{R_k^S, R_k^C \ \forall k\}$, $\bm{\theta}$, $\mathbf{z}$, $\{D_u \ \forall u\}$, $N_T$, $\epsilon_0$, $P_{max}$, $\gamma_{\textnormal{UL}}$, $B$, $\bm{\eta}^{i - 1}$, $\mathbf{P}^{i}$.
\begin{algorithmic}[1]
	\State Initialization:  $\bm{\eta}^{i, 0} = \bm{\eta}^{i - 1}$, $\epsilon = 1 \times 10^{-2}$, $j = 1$.
	\State Solve (\ref{pro2.2.1}), denoting the optimal solution as $(\delta^{*}_T, \bm{\eta}^*)$, set $\bm{\eta}^{i, j} =\bm{\eta}^*$, $\delta_T(\mathbf{P}^{i}, \bm{\eta}^{i, j}) = \delta^{*}_T$;
	\While{$|1 - \frac{\delta_T(\mathbf{P}^{i}, \bm{\eta}^{i, j - 1})}{\delta_T(\mathbf{P}^{i}, \bm{\eta}^{i, j})}| > \epsilon$}
	\State $j = j + 1$;
	\State Solve (\ref{pro2.2.1}), denoting the optimal solution as $(\delta^{*}_T, \mathbf{P}^*)$, set $\bm{\eta}^{i, j} =\bm{\eta}^*$, $\delta_T(\mathbf{P}^{i}, \bm{\eta}^{i, j}) = \delta^{*}_T$;
	\EndWhile
\end{algorithmic}
\hspace*{0.02in} {\bf Output:}
$\bm{\eta}^{i, j}, \delta_T(\mathbf{P}^{i}, \bm{\eta}^{i, j})$. \\
\end{algorithm}

\par{
After the problems in (\ref{pro2.1}) and (\ref{pro2.2}) are solved, the solution to (\ref{pro2}) can be iteratively derived by jointly using Algorithm 1 and Algorithm 2, according to block coordinate descent technique \cite{Liu2020}. Detailed steps of the proposed joint resource orchestration scheme are summarized in Algorithm 3. 
}
\addtolength{\topmargin}{0.05in}
\begin{algorithm}[t]
	\caption{Joint resource orchestration algorithm to solve (\ref{pro2})}
	\hspace*{0.02in} {\bf Input:} $K$, $M$, $U$, $R^L$, $\{R_k^S, R_k^C \ \forall k\}$, $\bm{\theta}$, $\mathbf{z}$, $\{D_u \ \forall u\}$, $N_T$, $\epsilon_0$, $P_{max}$, $\gamma_{\textnormal{UL}}$, $B$.
	\begin{algorithmic}[1]
		\State Initialization:  $\mathbf{P}^0 = 1 \times 10^{-3} \mathbf{U}_{U \times N_T}$, where the elements in $\mathbf{U}_{U \times N_T}$ are uniformly distributed random variables in the range of $[0, 1]$, $\epsilon = 1 \times 10^{-3}$, $i = 1$;
		\State Solve (\ref{pro2.2}) using Algorithm 2, denoting the optimal solution as $(\delta^{*}_T, \bm{\eta}^*)$, set $\bm{\eta}^{0} =\bm{\eta}^*$, $\delta_T(\mathbf{P}^{0}, \bm{\eta}^{0}) = \delta^{*}_T$;
		\State Solve (\ref{pro2.1}) using Algorithm 1, denoting the optimal solution as $(\delta^{*}_T, \mathbf{P}^*)$, set $\mathbf{P}^{i} =\mathbf{P}^*$, $\delta_T(\mathbf{P}^{i}, \bm{\eta}^{i - 1}) = \delta^{*}_T$;
		\State Solve (\ref{pro2.2}) using Algorithm 2, denoting the optimal solution as $(\delta^{*}_T, \bm{\eta}^*)$, set $\bm{\eta}^{i} =\bm{\eta}^*$, $\delta_T(\mathbf{P}^{i}, \bm{\eta}^{i}) = \delta^{*}_T$;
		\While{$|1 - \frac{\delta_T(\mathbf{P}^{i - 1}, \bm{\eta}^{i - 1})}{\delta_T(\mathbf{P}^{i}, \bm{\eta}^{i})}| > \epsilon$}
		\State $i = i + 1$;
		\State Solve (\ref{pro2.1}) using Algorithm 1, denoting the optimal solution as $(\delta^{*}_T, \mathbf{P}^*)$, set $\mathbf{P}^{i} =\mathbf{P}^*$, $\delta_T(\mathbf{P}^{i}, \bm{\eta}^{i - 1}) = \delta^{*}_T$;
		\State Solve (\ref{pro2.2}) using Algorithm 2, denoting the optimal solution as $(\delta^{*}_T, \bm{\eta}^*)$, set $\bm{\eta}^{i} =\bm{\eta}^*$, $\delta_T(\mathbf{P}^{i}, \bm{\eta}^{i}) = \delta^{*}_T$;
		\EndWhile
	\end{algorithmic}
	\hspace*{0.02in} {\bf Output:}
	$\mathbf{P}^{i}, \bm{\eta}^{i}, \delta_T(\mathbf{P}^{i}, \bm{\eta}^{i})$. \\
\end{algorithm}
\subsection{The Solution to (\ref{pro3})}
\par{
Comparing (\ref{pro3aa}) with (\ref{pro2aa}), we can give the following property.
\begin{property}
	The optimal solution to (\ref{pro3}) must satisfy $\eta_{u, t}^C = 0$ for $\forall u,t$.
\end{property}
\begin{IEEEproof}
It is observed that (\ref{pro3aa}) is uncorrelated with $\eta_{u, t}^C$, so that the values of $\eta_{u, t}^C$ will not influence the value of $\delta_T$ in (\ref{pro3aa}). Furthermore, if $\eta_{u, t}^C = 0$ is substituted into (\ref{pro3}), all constraints in (\ref{pro3b})--(\ref{pro3g}) can be satisfied. As a result, $\eta_{u, t}^C = 0$ always belong to the feasible region of (\ref{pro3}) $\forall u,t$, which gives the conclusion.
\end{IEEEproof}
Using Property 2, (\ref{pro3}) can be simplified to
\begin{subequations}\label{pro3.1}
	\begin{align}
	\min_{\mathbf{P}, \bm{\eta}, \delta_T} & \ \delta_T \label{pro3.1a} \\
	\nonumber s.t. \ & \sum_{t=1}^{N_{T}} \delta_{T} \left( \frac{\eta^L_{t}}{R^L} + \frac{\eta^{S}_{u, t}}{\sum_{k = 1}^{K} z_{u, k} \hat{R}^U_{u,k,t}(\mathbf{P})} \right.  \\
	& \left. + \frac{\eta^{S}_{u, t}}{\sum_{k = 1}^{K} z_{u, k} R_k^S}  \right)^{-1} \geq D_u \ \forall u \label{pro3.1b} \\
	& \sum_{u=1}^U \eta_{u, t}^{S} z_{u, k} \hat{R}^U_{u, k, t}(\mathbf{P}) \leq R^S_k \ \ \forall k,t \label{pro3.1c}\\
	& 0 \leq p_{u, t} \leq P_{max} \ \ \forall u,t \label{pro3.1d} \\
	& \eta_{t}^L + \eta_{u, t}^S = 1 \ \ \forall u,t \label{pro3.1e} \\
	& 0 \leq \eta_{t}^L,\eta_{u, t}^S \leq 1 \ \ \forall u,t \label{pro3.1f} \\
	& \delta_T \geq 2 \epsilon_0 \label{pro3.1g}
	\end{align}
\end{subequations}
which can be solved using Algorithm 1--3 with $\eta_{u, t}^C = 0$ $\forall u,t$.
}
\begin{remark}
	\textnormal{The most important difference between the solution to (\ref{pro2}) and that to (\ref{pro3}) is that, whether or not MEC can be used for communication and computing. Similar to the discussions in Remark 1, the communication and computing process may be accomplished in one segmentation of the process when $D_u$ is small, whereas a smaller communication and computing latency can be achieved if MEC is not used. The reason for this is that the communication and computing latency could be lower than $3 \epsilon_0$ if MEC is not used, but the latency must be at least $3 \epsilon_0$ if MEC is used, as shown by (\ref{pro2g}) and (\ref{pro3g}). Such phenomenon can also be observed after numerical results are derived.
	}
\end{remark}
\par{
	On the basis of Algorithm 1--3, a process-oriented joint resource orchestration scheme to solve (\ref{pro0.1}) is derived,  and it is summarized in Algorithm 4. Using Algorithm 4, the minimum overall communication and computing latency can be obtained.
}
\addtolength{\topmargin}{0.05in}
\begin{algorithm}[t]
	\caption{Proposed process-oriented joint resource orchestration algorithm}
		\hspace*{0.02in} {\bf Input:} $K$, $M$, $U$, $R^L$, $\{R_k^S, R_k^C \ \forall k\}$, $\bm{\theta}$, $\mathbf{z}$, $\{D_u \ \forall u\}$, $N_T^{s_1}$, $N_T^{s_2}$, $N_T^{s_3}$, $\epsilon_0$, $\epsilon_a$, $P_{max}$, $\gamma_{\textnormal{UL}}$, $B$.
	\begin{algorithmic}[1]
		\State Solve (\ref{pro1}) with $N_T = N_T^{s_1}$; then, $(\mathbf{P}^{s_1}, \mathbf{\eta}^{s_1})$ is derived by (\ref{pro1eqn1}) and $\delta^{s_1}_T$ is derived by (\ref{pro1eqn2}); 
		\State Solve (\ref{pro2}) with $N_T = N_T^{s_2}$ using Algorithm 1--3. Denoting the solution as $(\mathbf{P}^{s_2}, \mathbf{\eta}^{s_2})$; then $\delta^{s_2}_T$ is derived;
		\State Solve (\ref{pro3}) with $N_T = N_T^{s_3}$ using Algorithm 1--3. Denoting the solution as $(\mathbf{P}^{s_3}, \mathbf{\eta}^{s_3})$; then $\delta^{s_3}_T$ is derived;
		\State Calculate $T_{min} = \min\{N_T^{s_1}\delta^{s_1}_T, N_T^{s_2}\delta^{s_2}_T, N_T^{s_3}\delta^{s_3}_T\} + \epsilon_a$, where $(\mathbf{P}^{*}, \bm{\eta}^{*})$ is derived as the corresponding joint power allocation and data stream scheduling scheme;
	\end{algorithmic}
	\hspace*{0.02in} {\bf Output:}
	$T_{min}$, $\mathbf{P}^{*}$, $\bm{\eta}^{*}$. \\
\end{algorithm}

\subsection{Convergence Analysis}
\par{
In this section, we analyze the convergence of Algorithm 1--3. For the problem in (\ref{pro2.1.1}) during the $j$-th iteration step, we have
\begin{equation} \label{ineq1}
	\delta_T(\mathbf{P}^{i, j}, \bm{\eta}^{i - 1}) \leq \delta_T(\mathbf{P}^{i, j - 1}, \bm{\eta}^{i - 1})
\end{equation}
because both $\mathbf{P}^{i, j}$ and $\mathbf{P}^{i, j - 1}$ are feasible solutions to (\ref{pro2.1.1}) according to Theorem 1, and the minimum value of $\delta_T$ is achieved by $\mathbf{P}^{i, j}$. Thus, Algorithm 1 is guaranteed to converge according to (\ref{ineq1}), where we can derive $\mathbf{P}^{i, *}$ as the locally optimal solution which satisfies
\begin{equation} \label{lim1}
\delta_T(\mathbf{P}^{i, *}, \bm{\eta}^{i - 1}) \leq \delta_T(\mathbf{P}^{i, j}, \bm{\eta}^{i - 1}), \ \mathbf{P}^{i, j} \rightarrow \mathbf{P}^{i, *}, \ j \rightarrow \infty.
\end{equation}
Similarly, for the problem in (\ref{pro2.2.1}) at the $j$-th iteration step, we have
\begin{equation} \label{ineq2}
\delta_T(\mathbf{P}^{i}, \bm{\eta}^{i, j}) \leq \delta_T(\mathbf{P}^{i}, \bm{\eta}^{i, j - 1})
\end{equation}
based on Property 1, which indicates that Algorithm 2 is also guaranteed to converge. Therefore, we have $\bm{\eta}^{i, *}$ as the locally optimal solution which satisfies
\begin{equation} \label{lim2}
\delta_T(\mathbf{P}^{i}, \bm{\eta}^{i, *}) \leq \delta_T(\mathbf{P}^{i}, \bm{\eta}^{i, j}), \ \bm{\eta}^{i, j} \rightarrow \bm{\eta}^{i, *}, \ j \rightarrow \infty.
\end{equation}
According to (\ref{ineq1})--(\ref{lim2}), we can conclude that
\begin{equation}
\delta_T(\mathbf{P}^{i, *}, \bm{\eta}^{i, *}) \leq	\delta_T(\mathbf{P}^{i, *}, \bm{\eta}^{i - 1, *}) \leq \delta_T(\mathbf{P}^{i - 1, *}, \bm{\eta}^{i - 1, *})
\end{equation}
which shows that the objective function in (\ref{pro2a}) keeps decreasing when $i$ goes larger. Due to the constraints in (\ref{pro2aa})--(\ref{pro2g}), the objective function must have lower bound. As a result, the convergence of Algorithm 3 is proved, and the locally optimal solution to (\ref{pro2}) can be derived.
}

\section{Simulation Results and Discussions}

\par{
In this section, we evaluate the performance of proposed algorithms by simulation results. Parameters of the hierarchical NTN are set as $K = 7$, $M = 8$, $U = 56$, where $U$ IoT devices are divided into $K$ user groups and each device is associated with the nearest UAV. The positions of devices and UAVs are generated according to \cite{Mirahsan2015}, where the degree of user aggregation is set as $\beta = 0.5$. The minimum distance between UAVs is set as $d_{\textnormal{UAV}} = 30$ km, and the height of UAV swarm is set as $h_k = 3$ km $\forall k$. We assume that the data sizes of all IoT devices are the same, and this size is denoted as $D_u = D$ where we set $D = 1$ Gbits $\forall u$, and the maximum transmit power of each IoT device is set as $P_{max} = 2$~W \cite{Zhao2020}. For the satellite system, the data rate of device-satellite link is set as $R^{L} = 9.6$~kbits/s \cite{Dymond1988}, the maximum rate of UAV-satellite link is set as $R^{S}_{k} = 2$~Mbits/s $\forall k$ \cite{Peng2012}, and the overall propagation time of electromagnetic wave is set as $\epsilon_a = 240$~ms \cite{Luglio2014}. For UAV channel parameters, we set $m = 4.02$, $\Omega = 25 \times 10^{-3}$, $\eta_{\textnormal{LOS}} = 0.1$, $\eta_{\textnormal{NLOS}} = 21$, $f = 5.8$~GHz, $c = 3 \times 10^8$~m/s, $\lambda = c / f$, $d_0 = \lambda / 2$, $a = 5.0188$ and $b = 0.3511$ \cite{Chen2018, Khuwaja2018}. For the integrated communication and MEC system, we set $\zeta_u = 0.01$ $\forall u$ \cite{Sharma2018}, $\gamma_{\textnormal{UL}} = 0.1$, $B = 1$~MHz, $N_T^{s_1}=N_T^{s_2}=N_T^{s_3}=8$, $\sigma^2 = -114$~dBm \cite{Cao2016} and $\epsilon_0 = 500$~ms \cite{Luglio2014}\footnote{All of the parameters mentioned above are kept constant in the simulations unless otherwise specified.}. 
}
\begin{figure}[t] 
	\centering
	\includegraphics[width=3.8in]{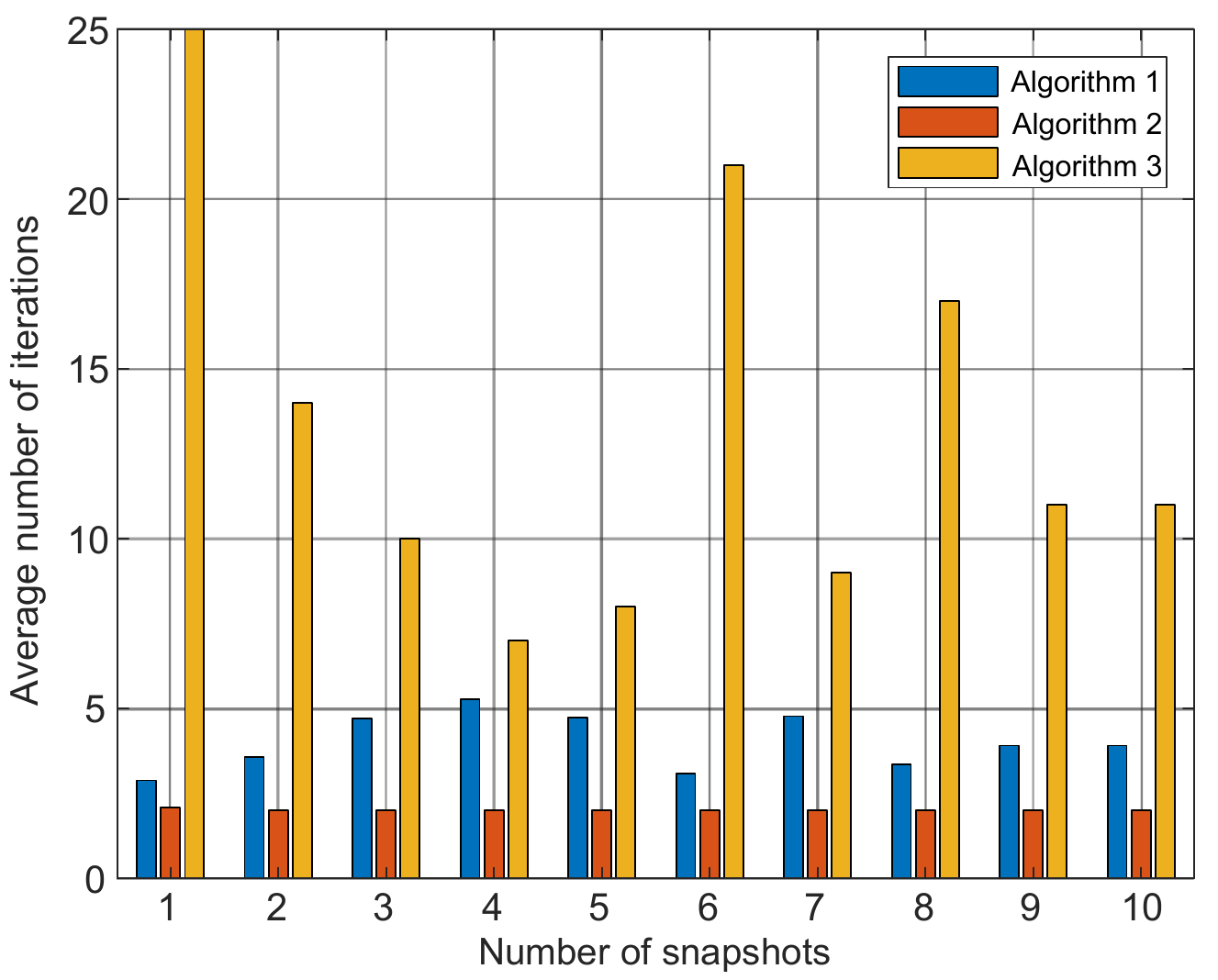}
	\caption{ Convergence performances of the proposed algorithms.}
	\label{figconverge}
\end{figure}
\par{
First, we evaluate the convergence performance of the proposed algorithms by numerical simulations, where we set $R^C_{k} = 10$ Mbits/s $\forall k$ \cite{Wang2020TVT}, and $N_T$ is randomly generated in each snapshot. As shown in Fig.~\ref{figconverge}, we can find that Algorithm 1 converges after approximately $5$ iterations, while Algorithm 2 only needs $2$ iterations to converge, showing that the data stream scheduling subproblem is nearly a convex problem. In addition, because Algorithm 3 requires higher precision than the other algorithms, more iterations are needed to converge. These results indicate that the proposed process-oriented scheme has much potential for use in practical systems.
}

\begin{figure}[t] 
	\centering
	\includegraphics[width=3.8in]{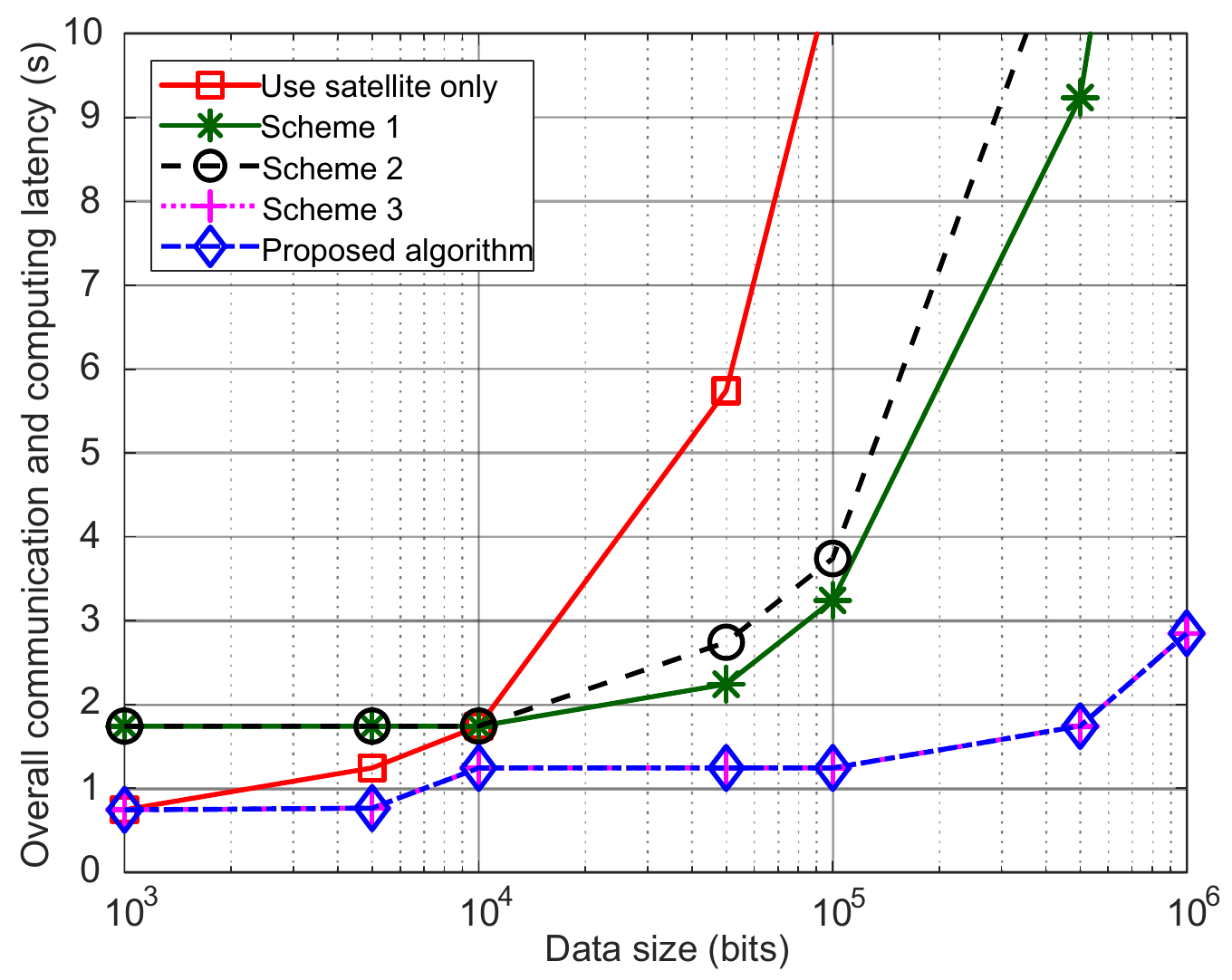}
	\caption{ Comparison between different algorithms when $D$ is small.}
	\label{figschemes1}
\end{figure}
\begin{figure}[t] 
	\centering
	\includegraphics[width=3.8in]{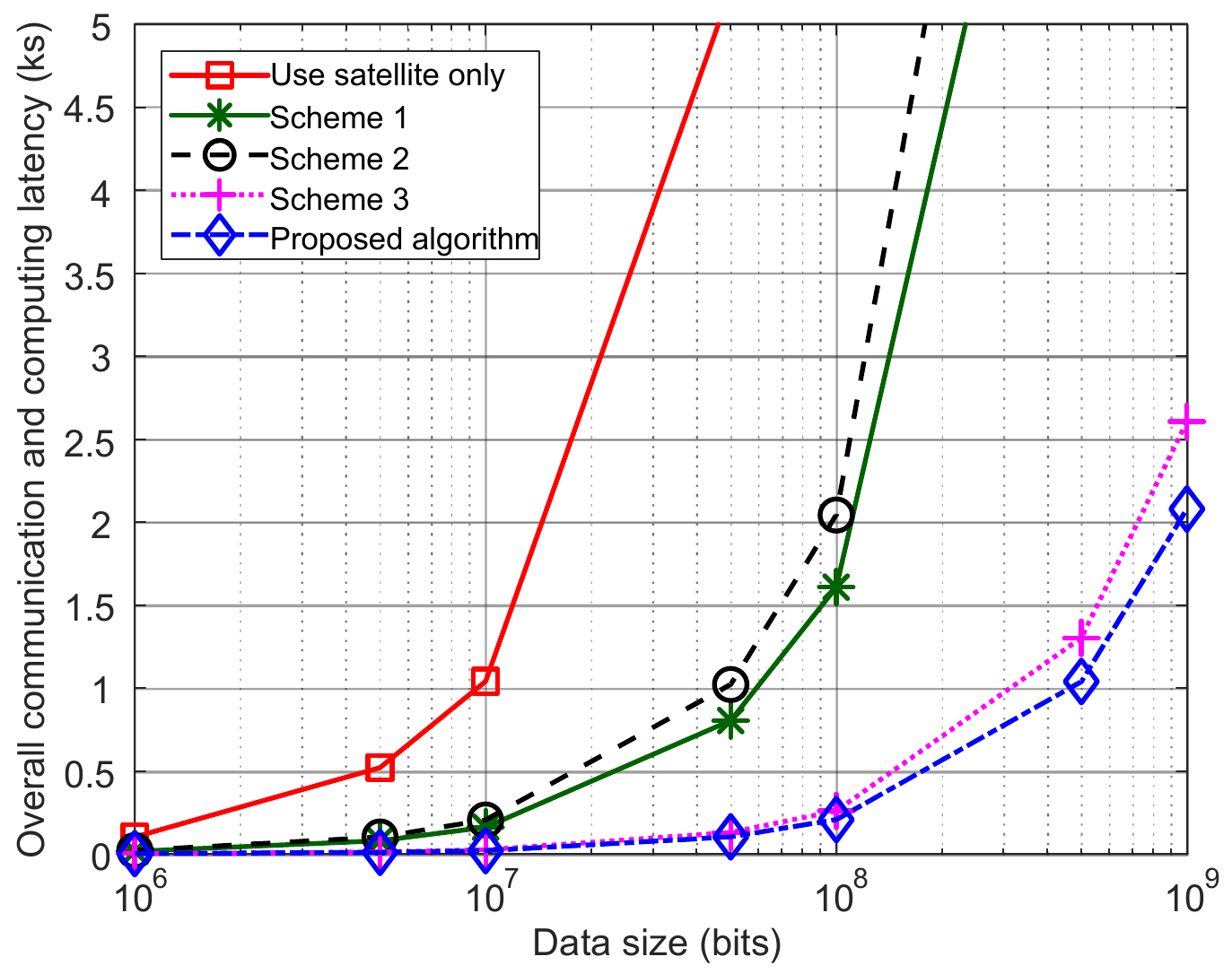}
	\caption{ Comparison between different algorithms when $D$ is large.}
	\label{figschemes2}
\end{figure}

\par{
Then, we compare the proposed algorithm with other schemes. A simple scheme is considered first, where we only use the satellite for communication. Furthermore, the following three schemes are also considered.
\begin{itemize}
	\item Scheme 1: We allocate the total bandwidth among multiple devices using the bandwidth allocation method proposed in \cite{Pan2020}, where the maximum transmission power is used, while a greedy data stream scheduling strategy is applied according to \cite{Cheng2019}.
	\item Scheme 2: The transmission power of each IoT device is set to be equal \cite{Liu2020}, where power backoff is used to satisfy the constraints of the data stream, while a greedy data stream scheduling strategy is applied according to \cite{Cheng2019}.
	\item Scheme 3: A simplified version of the proposed algorithm is used, where we assume that $N_T^{s_1}=N_T^{s_2}=N_T^{s_3}=1$ always holds.
\end{itemize}
In this simulation, we set $R^C_{k} = 6$ Mbits/s $\forall k$, and different algorithms are evaluated when the data sizes are varying. Observing Fig.~\ref{figschemes1} and Fig.~\ref{figschemes2}, we can find that the proposed algorithm has the best performance in comparison with other algorithms, but the performance gain varies with different data sizes. As illustrated in Fig.~\ref{figschemes1}, when $D$ is less than $1$ Mbits, the proposed algorithm has a similar performance to that of Scheme 3 because the communication and computing process can be accomplished in one segmentation of the process when the data size is small. In addition, a piecewise communication and computing latency pattern can be observed when the proposed algorithm is used because the design of the proposed scheme can adapt to varying data sizes. This phenomenon demonstrates the advantages of jointly designing power allocation and data stream scheduling schemes. Furthermore, the curves in Fig.~\ref{figschemes2} demonstrate that an approximately  $30$ percent performance gain is achieved by the proposed algorithm compared with Scheme 3, because $N_T$ and $\delta_T$ can be appropriately designed using Algorithm 1--4. To further clarify the insights of the proposed scheme, some typical values of communication and computing latency are demonstrated in Table~II, where the data sizes and update intervals are varying. The minimum communication and computing latencies with different data sizes are marked in bold type, and these demonstrate the phenomenon described in Remark 1 and Remark 2. It can also prove that the proposed scheme has the ability to adapt to different data sizes. From these results, we can conclude that it is beneficial to use the proposed process-oriented scheme in a hierarchical NTN with MEC. 
}
\begin{table} 
	\centering  
	\caption{Overall communication and computing latency with different data sizes}  
	\begin{tabular}{c|c|c|c}
		\toprule[1pt]
		\multirow{3}{*}{Data size} & \multicolumn{3}{|c}{Overall communication and computing latency derived by using the proposed scheme (Unit: s)}   \\
		\cline{2-4}
		& Only use satellite for   & Use satellite and UAVs for communication & Use satellite and UAVs for communication  \\
		& communication & without MEC & with MEC \\
		\hline
		$1$ Kbits & $\bf{0.74}$ & $1.24$ & $1.74$ \\
		\hline
		$10$ Kbits & $1.74$ & $\bf{1.24}$ & $1.74$ \\
		\hline
		$100$ Kbits & $10.74$ & $\bf{1.24}$ & $1.74$ \\
		\hline
		$1$ Mbits & $104.74$ & $5.24$ & $\bf{3.24}$ \\
		\hline
		$10$ Mbits & $1.04 \times 10^3$ & $46.24$ & $\bf{22.74}$\\
		\hline
		$100$ Mbits & $1.04 \times 10^4$ & $451.24$ & $\bf{213.24}$ \\
		\hline
		$1$ Gbits & $1.04 \times 10^5$ & $4.50 \times 10^3$& $\bf{2.12 \times 10^3}$ \\
		\bottomrule[1pt]
	\end{tabular}  
\end{table}  
\begin{figure}[t] 
	\centering
	\includegraphics[width=3.8in]{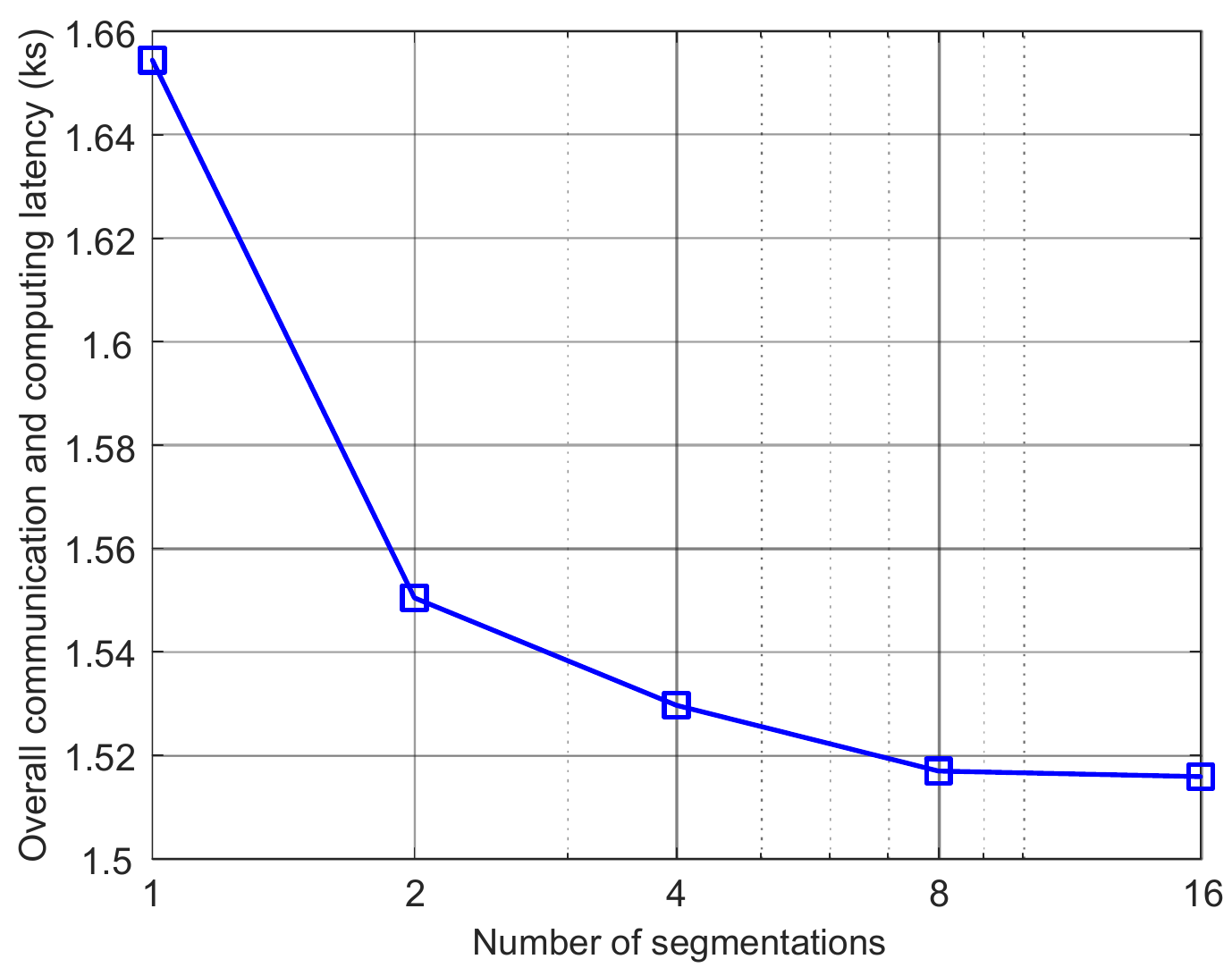}
	\caption{ The relationship between the minimum overall communication and computing latency and the segmentation numbers.}
	\label{fig_NT}
\end{figure}
\par{
In Fig.~\ref{fig_NT}, we evaluate the minimum overall latency derived by the proposed algorithms when the process is designed with different numbers of segmentations, where we set $R^C_k = 6$ Mbits/s $\forall k$ and $N_T^{s_1}=N_T^{s_2}=N_T^{s_3}=N_T$ as the segmentation number. On one hand, we can observe that the latency performance can be improved by dividing the process into more segmentations. On the other hand, the computational complexity is also increased with more segmentations, but the performance gain that we can acquire become less. These results indicate that the value of segmentation number should be appropriately selected in order to balance the computational complexity and the performance of the proposed algorithms. 
}

\begin{figure}[t] 
	\centering
	\includegraphics[width=3.8in]{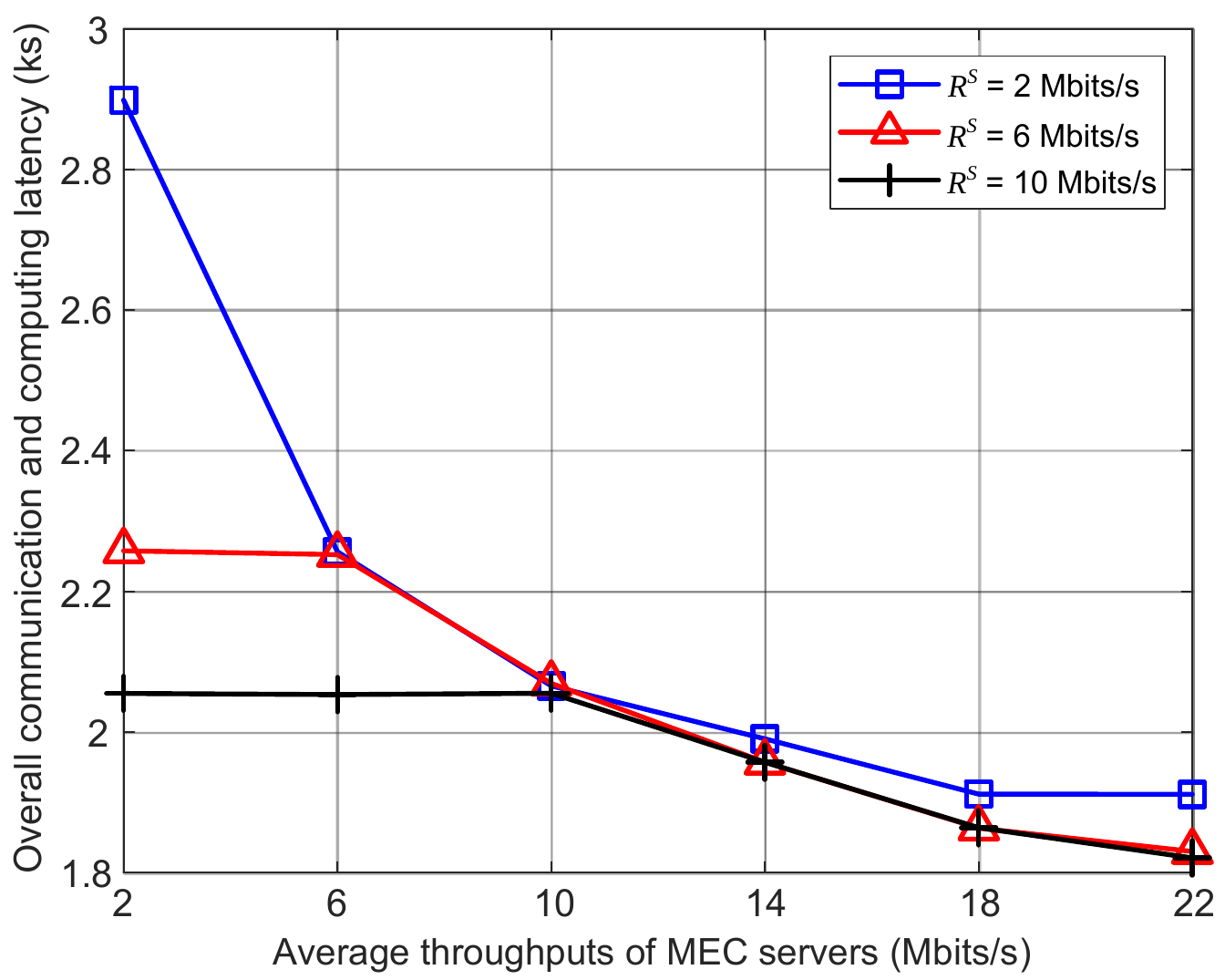}
	\caption{ The relationship between the overall communication and computing latency and average throughputs of MEC servers with varying data rates of UAV-satellite links.}
	\label{fig_RCRS}
\end{figure}
\par{
In Fig.~\ref{fig_RCRS}, we investigate the influence of UAV payload on  the latency performance of the proposed algorithm, where the average throughputs of MEC servers and the data rates of UAV-satellite links are varying. In this simulation, we set $R^S_{k} = R^S$ and $R^C_{k} = R^C$ $\forall k$. As illustrated in Fig. \ref{fig_RCRS}, the overall communication and computing latency is kept constant with respect to $R^{C}$ when $R^{C}$ is smaller than $R^{S}$because of the fact that most data would be scheduled to be sent back to the satellite through UAV-satellite links. Besides, we can also observe that the latency cannot be infinitely reduced by increasing $R^{C}$ because the limited data rates of device-UAV links may become the bottleneck in this case. These phenomena imply that the payload deployment of UAVs should be appropriately designed prior to the communication and computing process, as this could improve the efficiency of resource consumption regarding communications and MEC in a hierarchical NTN.
}

\begin{figure}[t] 
	\centering
	\includegraphics[width=3.8in]{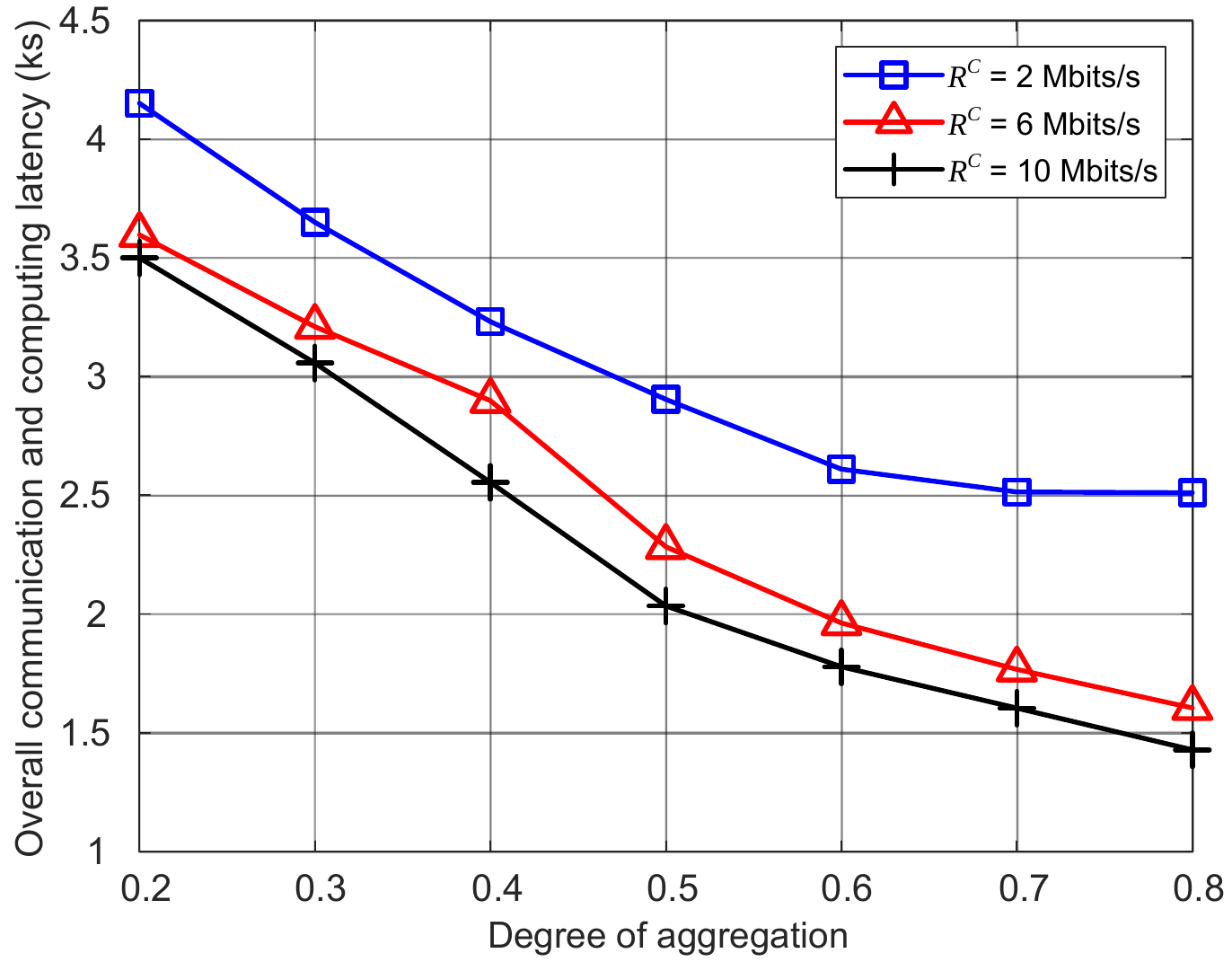}
	\caption{ The relationship between the overall communication and computing latency and degree of aggregation with different average throughputs of MEC servers.}
	\label{fig_aggregation}
\end{figure}

\par{
In Fig.~\ref{fig_aggregation}, we discuss the influence of user distribution on latency performance, where varying average throughputs of MEC servers are regarded. We set $R^C_{k} = R^C$ $\forall k$ in the simulation. From the curves, we can observe that decreased latencies can be achieved when $\beta$ increases because the proposed process-oriented scheme can reduce the interference between neighboring devices to some extent due to the fact that the diversity in time is fully used under the process-oriented framework. Specifically, when $\beta \geq 0.7$ and $R^C = 2$ Mbits/s, changing the degree of aggregation will not influence the latency performance, because limited throughput of MEC is the bottleneck for latency in this case. Moreover, it can be found that the gaps between curves decrease when $R^C$ continues to increase, and this can be illustrated by the phenomena shown in Fig.~\ref{fig_RCRS}.
}

\begin{figure}[t] 
	\centering
	\includegraphics[width=3.8in]{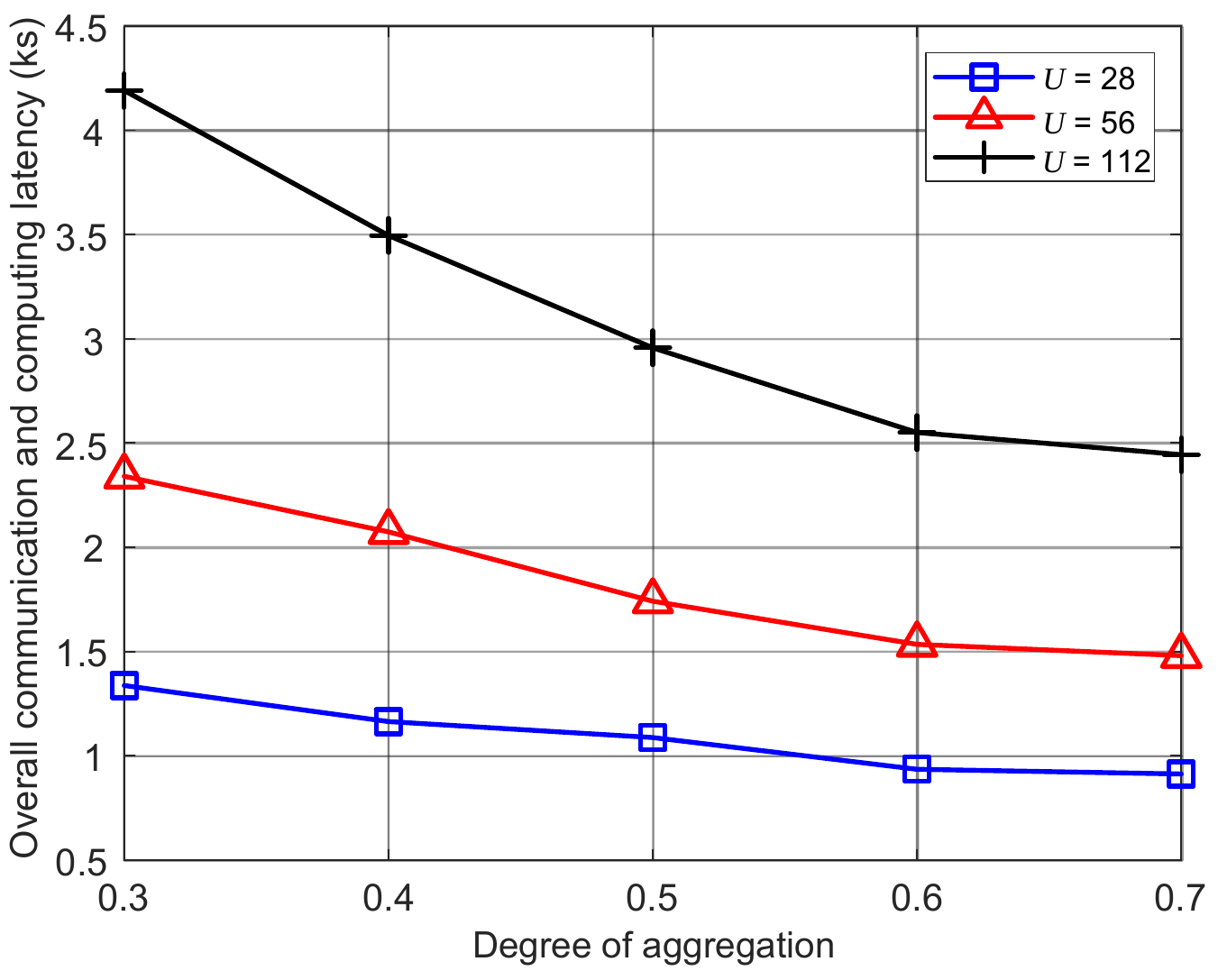}
	\caption{ The relationship between the overall communication and computing latency and degree of aggregation with different numbers of IoT devices.}
	\label{fig_Numbers}
\end{figure}
\par{
Furthermore, we evaluate the change of latency performance when different numbers of IoT devices are served in Fig.~\ref{fig_Numbers}. In the simulation, we set $R_k^C = 6$ Mbits/s $\forall k$ and $M = 16$. It is observed that the overall latency is larger when more IoT devices are served, because the resources of communication and MEC are more limited for more devices. Moreover, we can observe that the degree of aggregation has more influence on the latency performance with larger numbers of devices, for the reason that the interferences between IoT devices have more chances to be increased. These results indicate that the latency performance of MEC-empowered NTN is sensitive to user distribution, especially when the number of IoT devices is large in this network.
}

\begin{figure}[t] 
	\centering
	\includegraphics[width=3.8in]{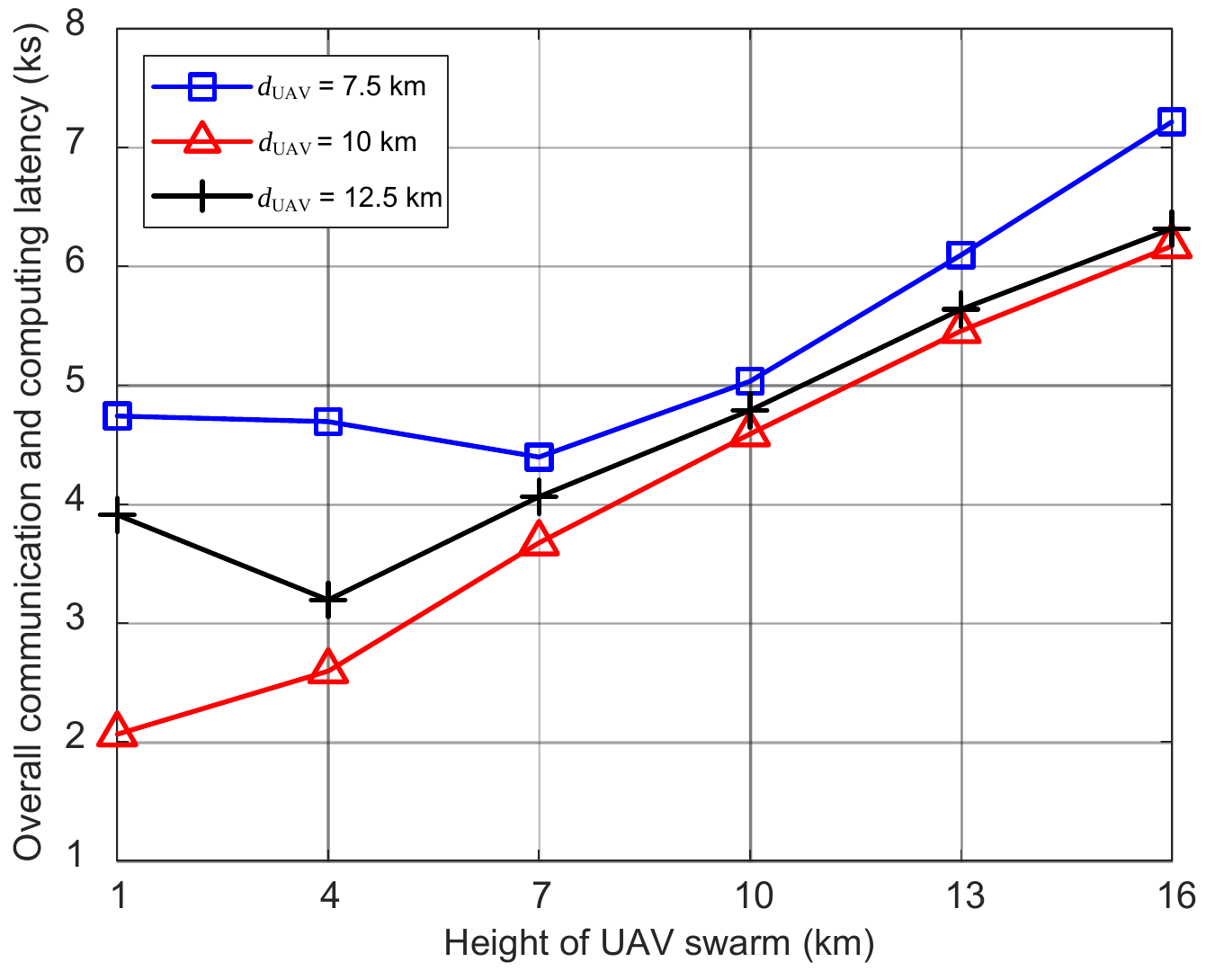}
	\caption{ The relationship between the overall communication and computing latency and height of UAV swarm, where the distance between different UAVs is varying.}
	\label{fig_range_high}
\end{figure}

\par{
Furthermore, we investigate the relationship between UAV positions and latency performance in Fig.~\ref{fig_range_high}, where we set $R^C_{k} = 10$ Mbits/s $\forall k$. As elaborated by the curves, it is observed that the overall communication and computing latency does not always monotonically increase with respect to the height of the UAV swarm. The reason is that although the path loss is larger with higher UAVs, the interference between multiple devices could decrease, showing that higher altitudes of UAVs may have positive effects on the latency performance of the algorithms. Such results indicate that the position and height of the UAV swarm should be appropriately designed in a hierarchical NTN to improve its latency performance.
}

\section{Conclusions}
\par{
In this paper, the design of an MEC-empowered NTN for wide-area time-sensitive IoT has been investigated. To jointly design the communication and MEC systems for hierarchically integrated satellite and UAVs, a process-oriented framework has been presented in a time-division manner. Under this framework, a latency minimization problem has been formulated using the large-scale CSI. Then, the problem can be transformed to a simplified form. After that, an approximation of the simplified problem has been derived. Then, based on the properties of the overall communication and computing efficiency function, the approximated problem has been decomposed into subproblems. Furthermore, an iterative algorithm has been proposed to solve these subproblems by jointly using block coordinate descent and successive convex approximation techniques. Based on the solutions to the subproblems, a process-oriented joint resource orchestration scheme has been proposed for the MEC-empowered NTN. Simulation results have demonstrated that the proposed process-oriented scheme has the best performance in comparison with those of other algorithms. In addition, it has been observed in simulations that the proposed process-oriented scheme can flexibly adapt to varying data sizes. These results have also indicated that the payload deployments of UAVs should be appropriately predesigned to improve the efficiency of resource use in the hierarchical NTN. Furthermore, the results have implied that it is advantageous to integrate NTN with MEC for wide-area time-sensitive IoT.
}

\begin{appendices}
	
\section{Justification of the Ergodic Rate Approximation}
\par{
According to the signal model in (\ref{signal}), for any given normalized detection vector which satisfies $||\mathbf{w}|| = 1$, the received signal of the $u$-th IoT device at the $t$-th time can be expressed as
\begin{equation} \label{B.1}
	\mathbf{w}^{H} \mathbf{y}_{u, k, t} = \mathbf{w}^{H} \mathbf{h}_{u, k, t} x_{u, t} + \sum_{v = 1, v \neq u}^{U} \mathbf{w}^{H} \mathbf{h}_{v, k, t} x_{v, t} + \mathbf{w}^{H} \mathbf{n}_{u, k, t}. \tag{A.1}
\end{equation}
In (\ref{B.1}), $\mathbf{h}_{u, k, t}$, $x_{u, t}$ and $\mathbf{n}_{u, k, t}$ can be regarded as independent variables. Therefore, when $U \rightarrow \infty$, we can apply Central Limit Theorem to (\ref{B.1}), so that $\sum_{v = 1, v \neq u}^{U} \mathbf{w}^{H} \mathbf{h}_{v, k, t} x_{v, t}$ can be regarded as a Gaussian random variable. Thus, we have 
\begin{align}  \label{B.2}
\nonumber	& \log_2 \left(1 + \frac{|\mathbf{w}^{H} \mathbf{h}_{u, k, t}|^2 p_{u, t}}{\sum_{v = 1, v \neq u}^{U} |\mathbf{w}^{H} \mathbf{h}_{v, k, t}|^2 p_{v, t} + \sigma^2} \right) \\
	& \approx \log_2 \left(1 + \frac{|\mathbf{w}^{H} \mathbf{h}_{u, k, t}|^2 p_{u, t}}{\sum_{v = 1, v \neq u}^{U} \mathbf{E} |\mathbf{w}^{H} \mathbf{h}_{v, k, t}|^2 p_{v, t} + \sigma^2} \right). \tag{A.2}
\end{align}
On the basis of (\ref{B.2}), the ergodic rate can be approximated by
\begin{align}  \label{B.3}
 \nonumber &\mathbf{E} \left\{ \log_2 \left(1 + \frac{|\mathbf{w}^{H} \mathbf{h}_{u, k, t}|^2 p_{u, t}}{\sum_{v = 1, v \neq u}^{U} |\mathbf{w}^{H} \mathbf{h}_{v, k, t}|^2 p_{v, t} + \sigma^2} \right) \right\} \\
& \approx \mathbf{E} \left\{ \log_2 \left(1 + \frac{|\mathbf{w}^{H} \mathbf{h}_{u, k, t}|^2 p_{u, t}}{\sum_{v = 1, v \neq u}^{U} \mathbf{E} |\mathbf{w}^{H} \mathbf{h}_{v, k, t}|^2 p_{v, t} + \sigma^2} \right) \right\}. \tag{A.3}
\end{align}
Furthermore, due to the generality of $\mathbf{w}$, (\ref{B.3}) is also right for MMSE detection vector, where $\mathbf{w}_{u, k, t}$ can vary with randomly changing $\mathbf{h}_{u, k, t}$ and  $\mathbf{w} = \frac{\mathbf{w}_{u, k, t}}{||\mathbf{w}_{u, k, t}||}$ can be substituted into (\ref{B.3}). Accordingly, we have
\begin{align}  \label{B.4}
\nonumber &  R^{U}_{u, k, t} (\mathbf{P}) \\
\nonumber & = \mathbf{E} \left\{ \log_2 \left(1 + \frac{|\frac{\mathbf{w}_{u, k, t}^{H}}{||\mathbf{w}_{u, k, t}||} \mathbf{h}_{u, k, t}|^2 p_{u, t}}{\sum_{v = 1, v \neq u}^{U} |\frac{\mathbf{w}_{u, k, t}^{H}}{||\mathbf{w}_{u, k, t}||} \mathbf{h}_{v, k, t}|^2 p_{v, t} +  \sigma^2} \right) \right\} \\
\nonumber & \approx \mathbf{E} \left\{ \log_2 \left(1 + \frac{|\frac{\mathbf{w}_{u, k, t}^{H}}{||\mathbf{w}_{u, k, t}||} \mathbf{h}_{u, k, t}|^2 p_{u, t}}{\sum_{v = 1, v \neq u}^{U} \mathbf{E} |\frac{\mathbf{w}_{u, k, t}^{H}}{||\mathbf{w}_{u, k, t}||} \mathbf{h}_{v, k, t}|^2 p_{v, t} + \sigma^2} \right) \right\} \\
& \leq  \log_2 \left(1 + \frac{\mathbf{E} |\frac{\mathbf{w}_{u, k, t}^{H}}{||\mathbf{w}_{u, k, t}||} \mathbf{h}_{u, k, t}|^2 p_{u, t}}{\sum_{v = 1, v \neq u}^{U} \mathbf{E} |\frac{\mathbf{w}_{u, k, t}^{H}}{||\mathbf{w}_{u, k, t}||} \mathbf{h}_{v, k, t}|^2 p_{v, t} + \sigma^2} \right) \tag{A.4}
\end{align}
where the last inequality holds because the log function is concave and Jensen inequality can be applied, which gives a justification of the approximation in (\ref{approrate}).

}

\section{Proof of Theorem 1}
\par{
We can derive the first-order and second-order derivative of $\hat{R}^U_{u, k, t}(\mathbf{P})$ as
\begin{align}
& \frac{\partial \hat{R}^U_{u, k, t}(\mathbf{P})}{\partial p_{u, t}} = \frac{(1-\gamma_{\textnormal{UL}}) B}{\textnormal{ln}2} \frac{\theta_{u, u, k}}{p_{u, t}\theta_{u, u, k} + I_{u, k}(\bar{\mathbf{p}}_{u, t}) + \sigma^2} \label{C.1} \tag{B.1} \\
\nonumber &\frac{\partial \hat{R}^U_{u, k, t}(\mathbf{P})}{\partial p_{v, t}} = \frac{(1-\gamma_{\textnormal{UL}}) B}{\textnormal{ln}2} \\
\nonumber & \times \frac{-p_{u, t} \theta_{u, u, k} \theta_{u, v, k}}{(p_{u, t}\theta_{u, u, k} + I_{u, k}(\bar{\mathbf{p}}_{u, t}) + \sigma^2)(I_{u, k}(\bar{\mathbf{p}}_{u, t}) + \sigma^2)} \ \forall v \neq u \label{C.2} \tag{B.2} \\
& \frac{\partial^2 \hat{R}^U_{u, k, t}(\mathbf{P})}{\partial p^2_{u, t}} = \frac{(1-\gamma_{\textnormal{UL}}) B}{\textnormal{ln}2} \frac{-\theta^2_{u, u, k} }{(p_{u, t}\theta_{u, u, k} + I_{u, k}(\bar{\mathbf{p}}_{u, t}) + \sigma^2)^2}  \label{C.3} \tag{B.3} \\
\nonumber &\frac{\partial^2 \hat{R}^U_{u, k, t}(\mathbf{P})}{\partial p^2_{v, t}} = \frac{(1-\gamma_{\textnormal{UL}}) B}{\textnormal{ln}2} \\
\nonumber &  \times \frac{p_{u, t} \theta_{u, u, k} \theta^2_{u, v, k} (p_{u, t} \theta_{u, u, k} + 2I_{u, k}(\bar{\mathbf{p}}_{u, t}) + 2 \sigma^2) }{(p_{u}\theta_{u, u, k} + I_{u, k}(\bar{\mathbf{p}}_{u, t}) + \sigma^2)^2(I_{u, k}(\bar{\mathbf{p}}_{u, t}) + \sigma^2)^2} \ \forall v \neq u \label{C.4} \tag{B.4} 
\end{align}
where we have
\begin{equation}
	\frac{\partial^2 \hat{R}^U_{u, k, t}(\mathbf{P})}{\partial p^2_{u, t}} \leq 0, \frac{\partial^2 \hat{R}^U_{u, k, t}(\mathbf{P})}{\partial p^2_{v, t}} \geq 0 \ \ \forall v \neq u \label{C.5} \tag{B.5}
\end{equation}
which shows that $\hat{R}^U_{u, k, t}(\mathbf{P})$ is concave with respect to $p_{u, t}$ and convex with respect to $\{p_{v, t} \ \ \forall v \neq u\}$. Then, we can reformulate $\hat{R}^U_{u, k, t}(\mathbf{P})$ to
\begin{align}
\nonumber	\hat{R}^U_{u, k, t}(\mathbf{P}) = & (1-\gamma_{\textnormal{UL}})B \{ \log_2(p_{u, t}\theta_{u, u, k} + I_{u, k}(\bar{\mathbf{p}}_{u, t}) + \sigma^2) \\
	&- \log_2(I_{u, k}(\bar{\mathbf{p}}_{u, t}) + \sigma^2) \}. \label{C.6} \tag{B.6}
\end{align}
Further by applying Taylor expansion to the first part and second part of (\ref{C.6}) at any given point $\mathbf{P}^0$ respectively, we have 
\begin{align}
\nonumber & -\log_2(I_{u, k}(\bar{\mathbf{p}}_{u, t}) + \sigma^2)  \geq - \textnormal{log}_2( I_{u, k}(\bar{\mathbf{p}}^0_{u, t}) + \sigma^2)\\
& - \frac{1}{\textnormal{ln}2 (I_{u, k}(\bar{\mathbf{p}}^0_{u, t}) + \sigma^2)}(I_{u, k}(\bar{\mathbf{p}}_{u, t})- I_{u, k}(\bar{\mathbf{p}}^0_{u, t}))  \label{C.7} \tag{B.7}
\end{align}
\begin{align}
\nonumber & \log_2(p_{u, t}\theta_{u, u, k} + I_{u, k}(\bar{\mathbf{p}}_{u, t}) + \sigma^2) \\
\nonumber & \leq  \textnormal{log}_2(p^0_{u, t} \theta_{u,u,k} + I_{u, k}(\bar{\mathbf{p}}_{u, t}^0) + \sigma^2) \\ 
\nonumber & +\frac{1}{\textnormal{ln}2 (p^0_{u, t} \theta_{u,u,k} + I_{u, k}(\bar{\mathbf{p}}_{u, t}^0) + \sigma^2)} [p_{u, t} \theta_{u,u,k} - p^0_{u, t} \theta_{u,u,k} \\
& + I_{u, k}(\bar{\mathbf{p}}_{u, t}) - I_{u, k}(\bar{\mathbf{p}}_{u, t}^0)] \label{C.8} \tag{B.8}
\end{align}
due to the fact that log function is concave. According to (\ref{C.6})--(\ref{C.8}), we can prove the inequalities in (\ref{th2.1}) and (\ref{th2.2}), which gives the conclusion of Theorem 1.
}
\section{Proof of Theorem 2}
\par{
Observing  (\ref{pro2.1.1}), we can find that the objective function in (\ref{pro2.1.1a}) and the constraints in (\ref{pro2.1.1c})--(\ref{pro2.1.1f}) are convex. For the simplification of notations, the constraints in (\ref{pro2.1.1b}) are reformulated to
\begin{equation}
f_u(\delta_T, \mathbf{P}^{i, j}) = \frac{D_u}{\delta_T} - \sum_{t = 1}^{N_T} g_u(\bar{R}^U_{u, t}(\mathbf{P}^{i, j} | \mathbf{P}^{i, j - 1})) \leq 0 \ \forall u \label{D.1} \tag{C.1}
\end{equation}
where
\begin{equation}
	g_u(x) = \frac{x}{ C_{u, t}(\bm{\eta}^{i - 1}) x + 1 - \eta^{L, i - 1}_t}. \label{D.2} \tag{C.2}
\end{equation}
The first-order and second-order derivative of $f_u(\delta_T, \mathbf{P}^{i, j})$ with respect to $\delta_T$ is derived as
\begin{align}
\nonumber	& \frac{\partial f_u(\delta_T, \mathbf{P}^{i, j})}{\partial \delta_T} = -\frac{D_u}{\delta^2_T} \leq 0 \\
	&\frac{\partial^2 f_u(\delta_T, \mathbf{P}^{i, j})}{\partial \delta_T^2} = \frac{2D_u}{\delta^3_T} \geq 0 \ \forall u \label{D.3} \tag{C.3}
\end{align}
which shows that $f_u(\delta_T, \mathbf{P}^{i, j})$ is monotonically decreasing and convex with respect to $\delta_T$ $\forall u$. Then, we can also give the first-order and second-order derivative of $g_u(x)$ as
\begin{align}
\nonumber & \frac{\partial g_u(x)}{\partial x} = \frac{1 - \eta^{L, i - 1}_t}{(C_{u, t}(\bm{\eta}^{i - 1}) x + 1 - \eta^{L, i - 1}_t)^2} \geq 0 \\
& \frac{\partial^2 g_u(x)}{\partial x^2} = -\frac{2C_{u, t}(\bm{\eta}^{i - 1}) (1 - \eta^{L, i - 1}_t)}{(C_{u, t}(\bm{\eta}^{i - 1}) x + 1 - \eta^{L, i - 1}_t)^3} \leq 0 \ \forall u \label{D.4} \tag{C.4}
\end{align}
which shows that $g_u(x)$ is monotonically increasing and concave with respect to $x$. According to \cite{Boyd2004}, $\bar{R}^U_{u, t}(\mathbf{P}^{i, j} | \mathbf{P}^{i, j - 1})$ is concave with respect to $\mathbf{P}^{i, j}$ $\forall u$, so that the composition function $g_u(\bar{R}^U_{u, t}(\mathbf{P}^{i, j} | \mathbf{P}^{i, j - 1}))$ is concave with respect to $\mathbf{P}^{i, j}$. As a result, $f_u(\delta_T, \mathbf{P}^{i, j})$ is convex with respect to the joint of $\delta_T$ and $\mathbf{P}^{i, j}$, which can guarantee the convexity of (\ref{pro2.1.1}).
}
\par{
Furthermore, denoting the optimal solution to (\ref{pro2.1.1}) as $(\delta_T^{*}, \mathbf{P}^{*})$, we have
\begin{align}
	\nonumber & \frac{D_u}{\delta^*_{T}} - \sum_{t=1}^{N_{T}}  \left( \frac{\eta^{L, i - 1}_{t}}{R^L} + \frac{\eta^{S, i - 1}_{u, t} + \eta^{C, i - 1}_{u, t}}{\sum_{k = 1}^{K} z_{u, k} \hat{R}^U_{u,k,t}(\mathbf{P}^*)} \right.  \\
	\nonumber & \left. + \frac{\eta^{S, i - 1}_{u, t} + \zeta_u \eta^{C, i - 1}_{u, t}}{\sum_{k = 1}^{K} z_{u, k} R_k^S} + \frac{\eta^{C, i - 1}_{u, t}}{\sum_{k = 1}^{K} z_{u, k} R_k^C} \right)^{-1} \\
	& \leq  \frac{D_u}{\delta^*_T} - \sum_{t = 1}^{N_T} g_u(\bar{R}^U_{u, t}(\mathbf{P}^{*} | \mathbf{P}^{i, j - 1})) \leq 0 \ \forall u \label{D.5} \tag{C.5}\\
\nonumber	& \sum_{u = 1}^U \eta_{u, t}^{C, i - 1} z_{u, k} 
	\hat{R}^U_{u,k,t}(\mathbf{P}^{*}) \leq \sum_{u = 1}^U \eta_{u, t}^{C, i - 1} z_{u, k} (1-\gamma_{\textnormal{UL}}) B   \\
	& \tilde{R}^U_{u, k}(p^{*}_{u, t}, I_{u, k}(\bar{\mathbf{p}}^{*}_{u, t})  | p^{i, j - 1}_{u, t}, I_{u, k}(\bar{\mathbf{p}}^{i, j - 1}_{u, t})) \leq  R^C_k \ \ \forall k,t \label{D.6} \tag{C.6} \\
\nonumber	& \sum_{u=1}^U (\eta_{u, t}^{S, i - 1} + \zeta_u \eta_{u, t}^{C, i - 1}) z_{u, k} \hat{R}^U_{u, k, t}(\mathbf{P}^{*}) \leq \sum_{u=1}^U (\eta_{u, t}^{S, i - 1}  \\
\nonumber	 & + \zeta_u \eta_{u, t}^{C, i - 1}) z_{u, k} (1-\gamma_{\textnormal{UL}}) B \tilde{R}^U_{u, k}(p^{*}_{u, t}, I_{u, k}(\bar{\mathbf{p}}^{*}_{u, t})  \\
 &| p^{i, j - 1}_{u, t}, I_{u, k}(\bar{\mathbf{p}}^{i, j - 1}_{u, t})) \leq R^S_k \ \ \forall k,t \label{D.7} \tag{C.7}\\
	& 0 \leq p^{*}_{u, t} \leq P_{max} \ \ \forall u,t \label{D.8} \tag{C.8} \\
	& \delta^*_T \geq 3 \epsilon_0 \label{D.9} \tag{C.9}
\end{align}
which demonstrates that $(\delta_T^{*}, \mathbf{P}^{*})$ also satisfies (\ref{pro2.1b})--(\ref{pro2.1f}). As a result, the conclusion of Theorem 2 is given.
}

\section{Proof of Property 1}
\par{
Denoting the optimal solution to (\ref{pro2.2.1}) as $(\delta_T^{*}, \bm{\eta}^{*})$, we have
\begin{align}
\nonumber s.t. \ & \frac{D_u}{\delta_T^{*}} - \sum_{t=1}^{N_{T}} \left( \frac{\eta^{L, *}_{t}}{R^L} + \frac{\eta^{S, *}_{u, t} + \eta^{C, *}_{u, t}}{\sum_{k = 1}^{K} z_{u, k} \hat{R}^U_{u,k,t}(\mathbf{P}^{i})} \right.  \\
\nonumber & \left. + \frac{\eta^{S, *}_{u, t} + \zeta_u \eta^{C, *}_{u, t}}{\sum_{k = 1}^{K} z_{u, k} R_k^S} + \frac{\eta^{C, *}_{u, t}}{\sum_{k = 1}^{K} z_{u, k} R_k^C} \right)^{-1} \\
\nonumber	& \leq \frac{D_u}{\delta^*_{T}} - \sum_{t=1}^{N_{T}}  \left\{ 
\frac{1}{C_{u, t}(\bm{\eta}^{i, j - 1}) + \frac{1 - \eta_t^{L, i, j - 1}}{\sum_{k = 1}^{K} z_{u, k} \hat{R}^U_{u,k,t}(\mathbf{P}^{i})}}
\right.  \\
\nonumber	& \left. - \frac{C_{u, t}(\bm{\eta}^{*}) - C_{u, t}(\bm{\eta}^{i, j - 1}) - \frac{\eta_t^{L, *} - \eta_t^{L, i, j - 1}}{\sum_{k = 1}^{K} z_{u, k} \hat{R}^U_{u,k,t}(\mathbf{P}^{i})} }{\left[C_{u, t}(\bm{\eta}^{i, j - 1}) + \frac{1 - \eta_t^{L, i, j - 1}}{\sum_{k = 1}^{K} z_{u, k} \hat{R}^U_{u,k,t}(\mathbf{P}^{i})}\right]^2} \right\} \\
&  \leq 0 \ \forall u \label{E.1} \tag{D.1} \\
& \sum_{u = 1}^U \eta_{u, t}^{C, *} z_{u, k} 
\hat{R}^U_{u,k,t}(\mathbf{P}^{i}) \leq R^C_k \ \ \forall k,t \label{E.2} \tag{D.2} \\
& \sum_{u=1}^U (\eta_{u, t}^{S, *} + \zeta_u \eta_{u, t}^{C, *}) z_{u, k} \hat{R}^U_{u, k, t}(\mathbf{P}^{i}) \leq R^S_k \ \ \forall k,t \label{E.3} \tag{D.3}\\
& \eta_{t}^{L, *} + \eta_{u, t}^{S, *} + \eta_{u, t}^{C, *} = 1 \ \ \forall u,t \label{E.4} \tag{D.4} \\
& 0 \leq \eta_{t}^{L, *},\eta_{u, t}^{S, *},\eta_{u, t}^{C, *} \leq 1 \ \ \forall u,t \label{E.5} \tag{D.5} \\
& \delta^*_T \geq 3 \epsilon_0 \label{E.6} \tag{D.6}
\end{align}
which shows that $(\delta_T^{*}, \bm{\eta}^{*})$ belongs to the feasible region of (\ref{pro2.2}). Thus, the conclusion of Property 1 is derived. 
}

\end{appendices}

\bibliographystyle{ieeetr}

\end{document}